\documentclass{aa}
\usepackage{psfig} 
\usepackage{lscape}
\usepackage{epsf}
\usepackage{graphics}

\def\ros{{\sl ROSAT }}

\begin{document}

\title{Soft X-ray properties of a spectroscopically selected sample of
interacting and isolated Seyfert galaxies \thanks{All overlays can be retrieved
via CDS anonymous ftp 130.79.128.5}}
\author{F. Pfefferkorn\inst{1}, Th. Boller\inst{1} and P. Rafanelli\inst{2}}     
\offprints{F. Pfefferkorn, pfefferk@mpe.mpg.de}
\institute{Max-Planck-Institut f\"ur extraterrestrische Physik, Postfach 1312, 85741 Garching, Germany 
           \and Department of Astronomy, University of Padova, Vicolo Osservatorio 5, 35122 Padova, Italy
	   } 
\date{Received: 11 October 2000; Accepted: 09 January 2001}

                                                            
\abstract{
We present a catalogue of ROSAT detected sources in the sample of spectroscopically selected Seyfert 1 and Seyfert 2 galaxies
of Rafanelli et al. (1995). The catalogue contains 102 Seyfert 1 and 36 Seyfert 2 galaxies. The identification is based on X-ray
contour maps overlaid on optical images taken from the Digitized Sky Survey. We have derived the basic spectral and timing properties
of the X-ray detected Seyfert galaxies.
For Seyfert 1 galaxies a strong correlation between photon index and X-ray luminosity is detected. We confirm the presence
of generally steeper X-ray continua in narrow-line Seyfert 1 galaxies (NLS1s) compared to broad-line 
Seyfert 1 galaxies. Seyfert 2 galaxies show photon indices similar to those of NLS1s.  
Whereas a tendency for an increasing X-ray luminosity with increasing interaction strength is found
for Seyfert 1 galaxies, such a correlation is not found for Seyfert 2 galaxies. 
For Seyfert 1 galaxies we found also a strong correlation for increasing far-infrared luminosity with increasing
interaction strength. Both NLS1s and Seyfert 2 galaxies show the highest values of far-infrared luminosity compared
to Seyfert 1 galaxies, suggesting that NLS1s and Seyfert 2 galaxies host strong (cirumnuclear) starformation.
For variable Seyfert galaxies we present the X-ray light curves obtained from the ROSAT All-Sky Survey and from ROSAT PSPC and HRI
pointed observations. Besides the expected strong short- and long-term X-ray variability in Seyfert 1 galaxies, we find indications for 
X-ray flux variations in Seyfert 2 galaxies.
}

\titlerunning{Soft X-ray properties of interacting and isolated Seyfert galaxies}
\authorrunning{F. Pfefferkorn et al.}
\maketitle

\section{Introduction}

Interaction between galaxies is considered to have a wide importance in triggering starburst and AGN activity,
which is thought to produce a rise in the luminosity.

Several studies of physical mechanisms which could
trigger a burst of strong star formation have been investigated by Barnes \& Hernquist (1991, 1996),
Jog \& Das (1992), Jog \& Solomon (1992) and Mihos \& Hernquist (1996).

Also many observations suggest that galactic encounters enhance star formation rates, as
demonstrated by studies of the optical (e.g., Kennicutt et al. 1987; Bushouse 1987), infrared
(e.g., Bernl\"ohr 1993; Telesco et al. 1993) and radio (e.g., Hummel 1981; Smith \& Kassim 1993)
emission from interacting systems. 
In the X-ray band, an enhanced starburst activity will result in an increase in the X-ray luminosity
due to the enhanced supernova rate. In addition, galaxy interaction might cause an increase in the
accretion rate onto the black hole, resulting in an additional increase in the X-ray luminosity. 

We have investigated the X-ray properties of a spectroscopically selected sample of Seyfert 1, 
NLS1 and Seyfert 2 galaxies with $ z<0.11, \; m_{\rm{v}} \le 15.5$ and $\delta \ge -23^\circ $ (Rafanelli et al.,1995).
The physical pairs have been selected on POSS plates using the following criteria: separation
between components $S < 3D_{\rm p}$, where $D_{\rm p}$ is the apparent major axis of the Seyfert galaxy, and magnitude-difference
between Seyfert galaxy and companion $\Delta m_{\rm v} = m_{\rm v,comp} - m_{\rm v,Seyfert} < 3$ (Rafanelli et al., 1995).
Approximately 30 \% of the \ros detected sources within the Rafanelli sample are Seyfert galaxies with separated physical
companions, 32 Seyfert 1 and 11 Seyfert 2 galaxies match the criteria mentioned above. Some of the isolated sources with
a disturbed morphology are probably merging galaxies or very close pairs, for which no separation $S$ is given by Rafanelli et. al.
(1995).  

The aim of this paper is to present the timing and spectral X-ray properties of Seyfert 1, NLS1 and Seyfert 2 galaxies. 
In Sect. 2 we describe the X-ray data analysis as well as the identification
process. The spectral features and timing results, including the relation between the
interaction strength $Q$ and the X-ray luminosity $L_{\rm X}$, are discussed in Sect. 3. 
The summary is given in Sect. 4. The database of X-ray detected Seyfert galaxies is quoted in the Appendix \ref{tables_light}.
The X-ray light curves of variable  Seyfert 1 galaxies are presented in Figs. \ref{fig_light1} to \ref{fig_light5}
of the Appendix \ref{tables_light}.


\section{ROSAT detected interacting and isolated Seyfert 1 \& 2 galaxies} 


\subsection{Identification procedure}

For each source of the Rafanelli sample we have searched for X-ray detections
within the \ros All-Sky Survey (RASS II catalogue; see Voges, Aschenbach, Boller et
al., 1999 for references) and public PSPC\footnote{Position Sensitive Proportional Counter} and
HRI\footnote{High Resolution Imager} pointed observations.
The sources have been identified on X-ray contour maps overlaid to optical images taken from the Palomar Digitized
Sky Survey (DSS). We have generated background subtracted contour maps for each selected \ros observation available for
inspection as postscript files at {\scriptsize http://wave.xray.mpe.mpg.de/publications/papers/2001}. These files can also be retrieved via
anonymous ftp from the address {\scriptsize ftp.xray.mpe.mpg.de} \footnote{\scriptsize
subdirectory: publications/papers/2001/interacting-xray-seyferts}.
Overlays with a dark background indicate pointed observations (HRI for HRI-data, without label for PSPC-data), whereas the grey
background indicate ROSAT All-Sky survey data (PSPC). The quality of the identification of the X-ray source with the
Seyfert galaxy is given in tables \ref{rafdats1.tab} and \ref{rafdats2.tab}, 1 and 2 indicate high and low reliability of the
identification, respectively.


\subsection{Data reduction of survey and pointed observations}

During the \ros All-Sky Survey the sources took from $\sim$ 5 s to $\sim$ 30 s to pass across the field of view of the PSPC detector
on each orbit ($\sim$ 96 min). The difference in crossing time depends on the scan range over the field of view, which covers a
circular area with a $57'.3$ radius.
We have only counted crossing times of individual sources through the PSPC detector, when the centroid source position was at least
$5'$ inside the detector. This prevents an underestimate of the source count rate due to the PSF. The resulting total exposure
times, obtained taking the sum of crossing times of a source, are in the range of $\sim$ 200 s up to $\sim$ 2000 s spread
over several days. For the \ros All-Sky Survey data reduction we have developed our own software modules, to take into
account the requirements for data reduction of the survey scan mode.

The data analysis of pointed \ros PSPC and HRI observations was performed with the standard
MIDAS/EXSAS software (Zimmermann et al. 1994). For this type of observation, the source is 
stationary centered on the detector. The exposure times
are in general significantly higher with respect to the survey observations
(see tables \ref{Xdats1.tab} and \ref{Xdats2.tab} in Appendix \ref{tables_light}).

\subsection{Spectral analysis}
\label{analysis}

For each source with more than 60 photons detected in the All-Sky Survey and for all sources within PSPC-pointed observations
we have performed a spectral analysis.
A power-law model 

\begin{equation}
f_{\rm E} dE \propto E^{\Gamma + 1} dE
\end{equation}

fitted to the spectral data yields the power-law parameters; neutral absorbing hydrogen column 
density $N_{\rm H_{fit}}$, photon index $\Gamma$ and the monochromatic flux at 1~keV.
The term $f_{\rm E} dE$ is the galaxy's energy flux between
the photon energies $E$ and $E + dE$.
The soft X-ray flux in the energy range 0.1--2.4 keV, corrected for absorption by neutral hydrogen,   
was calculated using the spectral fit parameters.
In the case of $N_{\rm H_{fit}} < N_{\rm H_{gal}}$, we have used the galactic
absorption column density $N_{\rm H_{gal}}$
(Dickey \& Lockman 1990).\\
For sources with $\le$ 60 counts in the survey observations and for HRI observations we have converted the 
mean count rates to the flux using a power-law model with photon index fixed to $\Gamma=-2.3$, which is
the typical value found for extragalactic objects with \ros, and the galactic
hydrogen column density $N_{\rm H_{gal}}$.\\
 
The soft X-ray flux was converted to luminosity using Eq. (7) of Schmidt \& Green (1986): 

\begin{equation}
L(E_1,E_2) = 4\pi (\frac{c}{H_{\rm o}})^2 C(z,\Gamma) {A(z)}^2 f(E_1,E_2)
\label{Lx}
\end{equation}

where a power-law spectrum is assumed in the energy range $(E_1,E_2)$, so that the redshift-dependent
functions $C(z)$ and $A(z)$ are given by:

\begin{equation}
C(z,\Gamma) = (1+z)^{\Gamma - 2}
\end{equation}

\begin{equation}
A(z) = 2[(1+z)-\sqrt{1+z}]
\end{equation}

We adopted for the cosmological deceleration parameter $q_{\rm o}=\frac{1}{2}$ and for the Hubble constant 
$H_{\rm o}=\rm 75\:km\:s^{-1}\:Mpc^{-1}$. The assumed photon index $\Gamma$ is given in tables \ref{Xdats1.tab}
 and \ref{Xdats2.tab} and the redshift in tables \ref{rafdats1.tab} and \ref{rafdats2.tab}.\\


\subsection{Timing analysis}

We have compared the survey and pointed count rates of individual sources in order to study the long-term variability
(0.5 up to 8.0 years). For all pointed observations and for the All-Sky Survey data with more than 60 source counts we have
produced the corresponding X-ray light curves (timescales up to a few days).

To produce the light curve of a survey source ($\ge$ 60 counts), we have defined a source
cell with radius of $5'$. The background was determined from a source-free cell with a radius of $5'$ located in
the scan direction through the centroid position of the source. 
This was necessary, as the effective exposure time depends on the position of the scan direction and the background level
may change for each time the path of source crosses the detector.

To take the \ros wobble for pointed observations into account we have used a minimum binsize of 400 s.
From the light curve we have computed the mean count rate of the Seyfert 1 and Seyfert 2 galaxies,
which are given in tables \ref{Xdats1.tab} and \ref{Xdats2.tab}, respectively.\\
The light curves of variable Seyfert 1 galaxies are given in Appendix \ref{tables_light}.

\section{Results}

In this section we present the X-ray properties of interacting and isolated Seyfert 1 \& 2 galaxies.
A description of the database is given in Subsect. \ref{catalogue}.
The spectral properties are listed in tables \ref{Xdats1.tab} and \ref{Xdats2.tab}, respectively.
The light curves for variable Seyfert 1 galaxies are shown in Figs. \ref{fig_light1} to \ref{fig_light5}.
The X-ray light curves of Seyfert 2 galaxies are given in Subsect. \ref{varia_s2_sec}. 
Relations between $\Gamma$, $L_{\rm X}$, ($L_{\rm fir}$), $Q$ and the Seyfert type are
presented in Sect. \ref{relations}.
The correlation of the interaction strength $Q$ and the X-ray luminosity
$L_{\rm X}$ is described in Subsect. \ref{interaction}.


\subsection{The catalogue}
\label{catalogue}

We have detected 91 out of 99 Seyfert 1 and 47 out of 98 Seyfert 2 galaxies of the Rafanelli sample in the \ros X-ray band
in pointed and/or All-Sky Survey observations. We have performed spectral analysis of the \ros PSPC observations for sources
with more than 60 counts s$^{-1}$. The timing analysis has been performed for all sources. In this paper we only show light curves
of sources with significant X-ray variability. Spectral information from the survey data could be obtained for 59
Seyfert 1 galaxies and only for one Seyfert 2 galaxy. 
The optical and X-ray properties of Seyfert 1 \& 2 galaxies of this sample are listed in tables \ref{rafdats1.tab} \&
\ref{rafdats2.tab}.
The tables quote the name of the Seyfert galaxy (column 2), ROSAT name (column
3) , redshift $z$ (column 4), diameter of the Seyfert galaxy $D_{\rm p}$ (column 5),
diameter of the companion galaxy $D_{\rm c}$ (column 6), separation between the components $S$ (column 7), dimensionless gravitational
interaction strength $Q$ (column 8) (for description see Sect. \ref{interaction}) and the apparent visual magnitude $V$ (column 9).
The values of $z$ and $V$ are taken from Veron-Cetty \& Veron catalogue (1991)
and the units of $D_{\rm p}$, $D_{\rm c}$ and $S$ are mm (POSS plates)
with a scale $\sim 13.4''/\rm mm$ (Rafanelli et al., 1995). Columns 10 and 11 show the quality of the X-ray identifications both in 
the pointing and the survey observations, the identifications labeled either with 1 or 2, 1 and 2 indicating high and lower degree of
reliability, respectively.
In the last column we have listed the classification of the Seyfert type (Sy1.0, Sy1.2, Sy1.5, Sy1.8, Sy1.9, Sy2.0) taken from the
{\em Catalogue of Seyfert Galaxies} (Lipovetski et al., 1987). We have modified the Rafanelli et al. conventions 
(S1 = Sy1.0 + Sy1.2 + Sy1.5 and S2 = Sy1.8 + Sy1.9 + Sy2.0) to S1 = Sy1.0 + Sy1.2 + Sy1.5 + Sy1.8 + Sy1.9 and S2 = S2.0.
The classifications marked by a hash (\#) indicate Narrow Line Seyfert 1 galaxies (NLS1)
(e.g. Osterbrock \& Pogge; Boller, Brandt \& Fink 1996; Grupe 1996).\\

In Appendix \ref{tables_light} the X-ray properties obtained from the timing and spectral analysis are listed in
tables \ref{Xdats1.tab} and \ref{Xdats2.tab}. Columns 2 and 3 contain the \ros position.
We mostly give the centroid source position from the pointed observation with the higher exposure times.
The columns 4 and 5 list the count rates, columns 6 and 7 the corresponding exposure times,
columns 8 and 9 the fluxes and columns 10 and 11 the luminosities of the sources detected in \ros
pointing and survey observations, respectively. The survey count rates were taken from the RASS II catalogue and the pointing
count rates were computed from the light curves of the sources. 
The apices $p$ and $h$ in column 4 indicate that the source data are taken from a PSPC or HRI observation.
In the columns 8 and 9 we apply the apices $f$ and $c$ to mark the data produced by spectral fit or by count rates.
This specification applies also for columns 10 and 11.
The Galactic column density is given in column 12 (Dickey \& Lockman, 1990), while the column density obtained from the spectral fit
is given in column 13. The other spectral fit parameters, namely the monochromatic flux at 1keV and the photon index are also
given in columns 14 and 15, respectively.
The value $\Gamma=-2.3$ was used, if no reliable spectral fit could be obtained. When spectral information was available
from the survey as well as from the pointed data, we quote the results from the pointed observations.\\

In most cases for optically separated close pairs, we detected in the X-ray band an unresolved single source (see overlays\footnote
{overlays are available at\\ http://wave.xray.mpe.mpg.de/publications/papers/2001}).
The results of these spectral fits are listed in tables \ref{Xdats1.tab} and \ref{Xdats2.tab}. When we detected two separate
X-ray components, we created two spectra and we show the sum of the count rates, fluxes and luminosities and the single
fit parameters ($N_{\rm H}$, $f_{\rm 1keV}$, $\Gamma$) of the Seyfert galaxy in the tables.


\subsection{Relations between $\Gamma,L_{\rm X},L_{\rm fir},Q$ and Seyfert type}
\label{relations}

In this section we present the spectral properties of Seyfert 1 and Seyfert 2 galaxies in terms of the relations between 
the photon index $\Gamma$, the interaction strength $Q$, the X-ray luminosity
$L_{\rm X}$, the far-infrared luminosity $L_{\rm fir}$
and the Seyfert type.

\subsubsection{Relations with $\Gamma$}

In Fig. 1 we have correlated the photon index obtained from the power-law fit with the X-ray luminosity. Different
subtypes of Seyfert 1 galaxies are marked with different labels. For low-luminosity Seyfert 1's, below about
$\rm 10^{42}\ erg\ s^{-1}$, where a significant contribution from the starburst is expected to contribute
to the total luminosity, there is no clear trend between $ \Gamma_{\rm fit}$
and $L_{\rm X}$.
However, for 'normal' Seyfert 1 type galaxies, a clear trend of a steepening of the X-ray spectrum with
increasing X-ray luminosity is detected.  A possible explanation for this effect might be a shifted
and strengthened accretion disk spectrum in high-luminosity Seyferts. This is expected, as the high
X-ray luminosity is most probably related to the accretion rate and/or the
black hole mass (Frank, King, Raine, 1985).
When fitting a simple-power law to the spectral data in the
ROSAT energy band, steeper values for the photon index are expected to arise in the high luminosity Seyfert 1 galaxies.
Another well-known effect is also present in Fig. 1, i.e. the steeper X-ray continua of NLS1s compared to broad-line Seyfert 1
 galaxies (Boller, Brandt \& Fink 1996).

\begin{figure}[h]
\begin{center}
\centerline{\psfig{figure=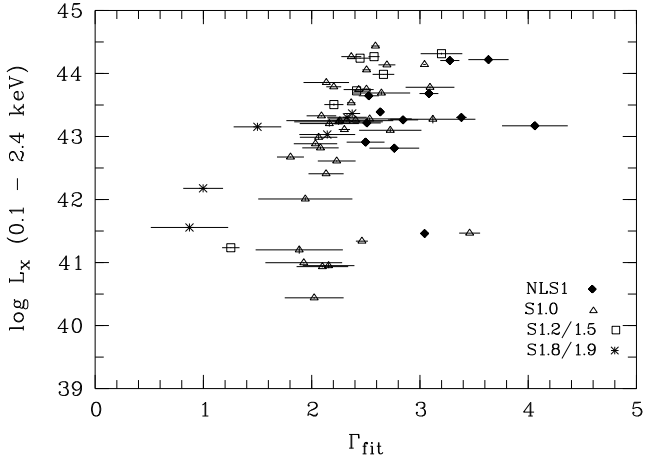,width=8.8cm,bbllx=15mm,bblly=15mm,bburx=200mm,bbury=275mm,angle=-90,clip=}}
\caption[]{
Relation between the X-ray luminosity $L_{\rm X} \; \rm [erg\,s^{-1}]$ and the photon index for Seyfert 1 galaxies. The different subtypes of Seyfert 1
galaxies (NLS1s, Seyfert 1.x) are plotted with different symbols. For Seyfert 1 galaxies above a X-ray luminosity
of about $\rm 10^{42}\ erg\ s^{-1}$ we find a strong trend of an increasing photon index with increasing X-ray luminosity.
}
\label{fig_Lx_gamma_s1}
\vspace{-0.5cm}
\end{center}
\end{figure}

For Seyfert 2 galaxies, we found no significant trend for an increasing photon index with 
increasing X-ray luminosity. Higher sensitivity measurements, e.g. with XMM-Newton,
are necessary to search for a correlation between the photon index and the X-ray luminosity for Seyfert 2 galaxies.

\begin{figure}[h]
\begin{center}
\centerline{\psfig{figure=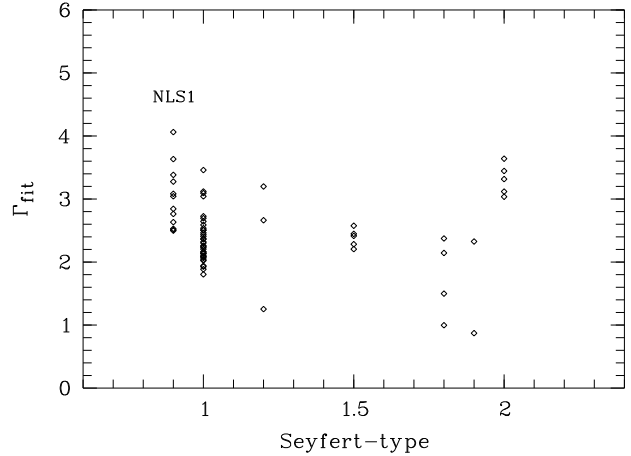,width=8.8cm,bbllx=15mm,bblly=15mm,bburx=200mm,bbury=275mm,angle=-90,clip=}}
\caption[]{Distribution of photon indices, obtained from the spectral fits, versus the Seyfert type (we have plotted
NLS1s at an x-axis value of 0.9). The highest values in the photon indices are found for NLS1s and for
Seyfert 2 galaxies.
}
\label{fig_gamma_type_s1}
\vspace{-0.5cm}
\end{center}
\end{figure}

In Fig. \ref{fig_gamma_type_s1} we show the distribution of photon indices, obtained from the spectral fits, versus the Seyfert type 
(we have plotted NLS1s at an x-axis value of 0.9). Only Seyfert galaxies with errors in the photon index smaller than
0.5 have been included. As expected, NLS1s show the largest values of the photon index, compared to Seyfert 1 galaxies.
Seyfert 2 galaxies show similar steep X-ray continua compared to NLS1s.
The Seyfert 1 galaxies show similar values of the photon indices as given by Walter \& Fink (1993).

\subsubsection{Interaction strength $Q$ - luminosity $L_{\rm X}$, $L_{\rm fir} $ relations}
\label{interaction}

In order to derive the interaction strength $Q$ we concentrate on the tidal force per unit mass produced
by a companion on a primary galaxy, which is proportional to $M_{\rm c} \cdot
R^{-3}$. $M_{\rm c}$ is the
mass of the companion and $R$ is its distance from the center of the primary galaxy. In most
cases, $M_{\rm c}$ and the absolute value of $R$ are unknown. Instead, these parameters are related to the dimensions
of the pair. Rubin et al. (1982) describe the dependence of the mass $M$ of a galaxy on the size
of its major axis $d$ as $M \propto d^{\gamma}$ and we use $\gamma = 1.5$ (Dahari, 1984). If we
use the apparent diameter of the primary galaxy $D_{\rm p}$ as a scaling factor, we obtain: 

\begin{equation}
M_{\rm c} \propto (D_{\rm c}/D_{\rm p})^{1.5} \qquad and \qquad R \propto S/D_{\rm p}
\end{equation}

Using these relations we get as dimensionless gravitational interaction strength $Q$:

\begin{equation}
\frac{M_{\rm c}}{R^3} \propto \frac{(D_{\rm c} \cdot D_{\rm p})^{1.5}}{S^3} \equiv Q
\end{equation}

This parameter is obviously large for close and relatively large companions.\\

\begin{figure}[h]
\begin{center}
\centerline{\psfig{figure=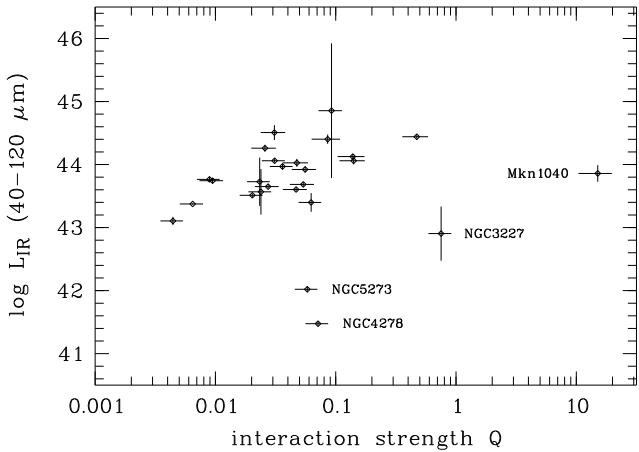,width=8.8cm,bbllx=15mm,bblly=15mm,bburx=200mm,bbury=275mm,angle=-90,clip=}}
\caption[]{Dimensionless gravitational interaction strength $Q$
vs. far-infrared luminosity $L_{\rm fir}\;\rm [erg\,s^{-1}]$ for Seyfert 1.x galaxies. The plot
suggests an increase in luminosity for increasing values of the interaction strength.}
\label{fig_Lir_Q_s1}
\vspace{-1.0cm}
\end{center}
\end{figure}

\begin{figure}[h]
\begin{center}
\centerline{\psfig{figure=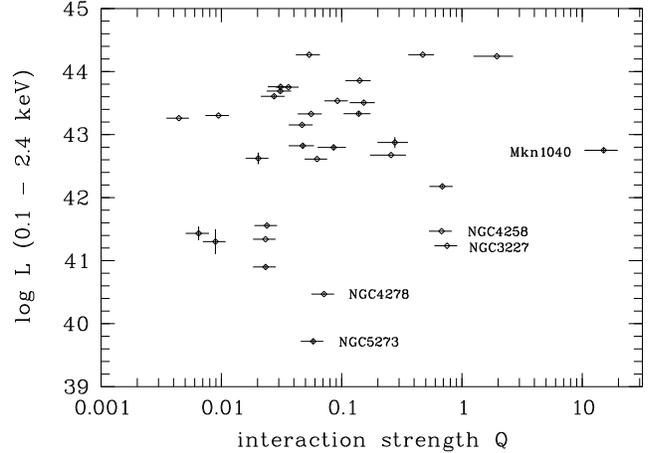,width=8.8cm,bbllx=15mm,bblly=15mm,bburx=200mm,bbury=275mm,angle=-90,clip=}}
\caption[]{Dimensionless gravitational interaction strength $Q$ vs. soft X-ray
luminosity $L_{\rm X}\;\rm [erg\,s^{-1}]$ for Seyfert 1.x galaxies. The plot
suggests an increase in luminosity for increasing values of the interaction strength. The marked objects with low luminosities are
probably absorbed sources with emission from the circumnuclear starbursts and scattered radiation from the nuclei.}
\label{fig_Lx_Q_s1}
\vspace{-1.0cm}
\end{center}
\end{figure}

Fig. \ref{fig_Lir_Q_s1} shows the interaction strength $Q$ vs. the far-infrared
luminosity $L_{\rm fir}$ for Seyfert 1
galaxies. For Seyfert 1 galaxies, the far-infrared luminosity increases with interaction strength.
The low-luminosity Seyfert 1 galaxies NGC~5273, NGC~4278, NGC~3227, (NGC~4258
is not detected in the IRAS Faint Source Catalogue) 
also show the trend of increasing X-ray and far-infrared luminosity with interaction strength Q. 
We speculate a large interaction strength of $Q=15.1\pm4.4$ for the galaxy
Mkn~1040 causes the relatively high X-ray (see fig. \ref{fig_Lx_Q_s1}) and far-infrared
luminosity and that the galaxy belongs to the low-luminosity population discussed above.
To test the correlation we have calculated the linear-correlation coefficient $r$
and the probability $P(r,N)$ for a linear correlation.
For the high-luminosity objects we obtain $r=0.3271$ and $P(r,22)\approx85\% \;(\sim1.5\sigma)$,
whereas we got for the low-luminosity Seyferts $r=0.8478$ and $P(r,4)\approx85\%\;(\sim1.5\sigma)$.

Fig. \ref{fig_Lx_Q_s1} shows the interaction strength $Q$ vs. the soft X-ray
luminosity $L_{\rm X}$ for Seyfert 1 galaxies.
For Seyfert 1 galaxies with $L_{\rm X} > 10^{42}$ there is a tendency for a luminosity increase with increasing interaction
strength. The labelled sources refer to the low-luminosity Seyfert 1 galaxies in our sample. As discussed for the
relation between the far-infrared luminosity and the interaction strength Q, both the low-luminosity Seyfert 1's
and the high-luminosity Seyfert 1's increase in X-ray luminosity when the interaction strength Q is increased.
The high spread of the distribution is quite likely produced by an overlap of effects from starburst and AGN.
In the case of infrared luminosity (fig. \ref{fig_Lir_Q_s1}) only the starburst play a role.
The linear-correlation test resulted in $r=0.2880$ and $P(r,27)\approx85\%\;(\sim1.5\sigma)$ for
the high-luminosity Seyferts and in $r=0.8233$ and $P(r,5)\approx92\%\;(\sim1.7\sigma)$ for the
low-luminosity objects.

The data points in Fig. \ref{fig_Lx_Q_s1} also include ROSAT pointed observations and Seyfert 1.8 and 1.9 galaxies, 
which complete the relation between $L_{\rm X}$ and $Q$ first discussed by Rafanelli et al. (1997).
For Seyfert 2 galaxies we found no correlation between the soft X-ray luminosity and the interaction strength. In Sect. \ref{discussion}
we discuss the problems in determining precisely the X-ray properties of obscured Seyfert 2 galaxies.


\subsubsection{Far-infrared relations}
\label{infrared}

In order to estimate the starburst activity we have investigated the
far-infrared luminosity $L_{\rm fir}$ using the far-infrared fluxes
$f_{\rm fir}$ at 60 and 100 $\rm \mu m$ from the IRAS Faint Source Catalogue. The total far-infrared fluxes
$f_{\rm fir}$ ($40 - 120 \rm \mu m$) were computed following Helou (1985) from
the IRAS 60 $\rm \mu m$ and 100 $\rm \mu m$ band fluxes:

\begin{equation}
f_{\rm fir} = 1.26 \cdot 10^{-11} (2.58 f_{60} + f_{100}) \;\rm erg\:cm^{-2}\,s^{-1} 
\label{F_ir}
\end{equation}

where $f_{60}$ and $f_{100}$ are given in Jansky. The far-infrared fluxes were converted to luminosities using Eq. \ref{Lx} in
Sect. \ref{analysis}. For the photon index we assumed $\Gamma = 1.5$.
  
\begin{figure}[h]
\begin{center}
\centerline{\psfig{figure=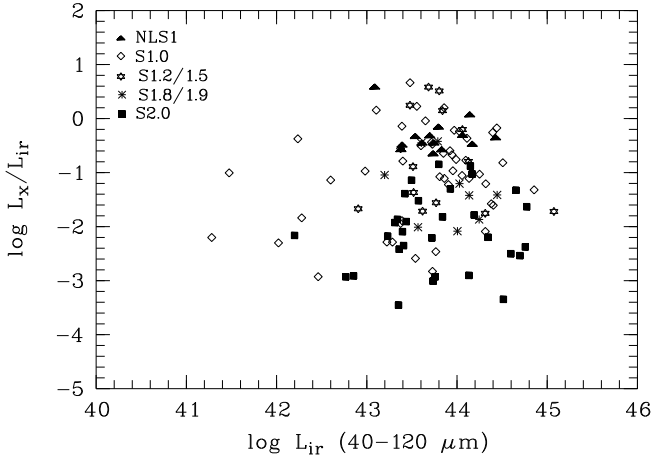,width=8.8cm,bbllx=15mm,bblly=15mm,bburx=200mm,bbury=275mm,angle=-90,clip=}}
\caption[]{Soft X-ray to far-infrared ($40-120 \rm \mu m$) luminosity ratio
$L_{\rm X}/L_{\rm fir}$ vs. far-infrared luminosity $L_{\rm fir}\;\rm [erg\,s^{-1}]$
for all types of Seyfert galaxies.}
\label{fig_Lx/Lir_Lir}
\vspace{-0.5cm}
\end{center}
\end{figure}

The ratio between far-infrared and soft X-ray luminosity $L_{\rm X}/L_{\rm
fir}$ and its dependence on the far-infrared luminosity $L_{\rm fir}$ is shown in fig.
\ref{fig_Lx/Lir_Lir} for all Seyfert types. Interestingly, NLS1 galaxies show a similar distribution of the far-infrared luminosity
as the Seyfert 2 galaxies. The far-infrared luminosity distribution is significantly different for Seyfert 1 galaxies compared
to NLS1 and Seyfert 2's.

\begin{figure}[h]
\begin{center}
\centerline{\psfig{figure=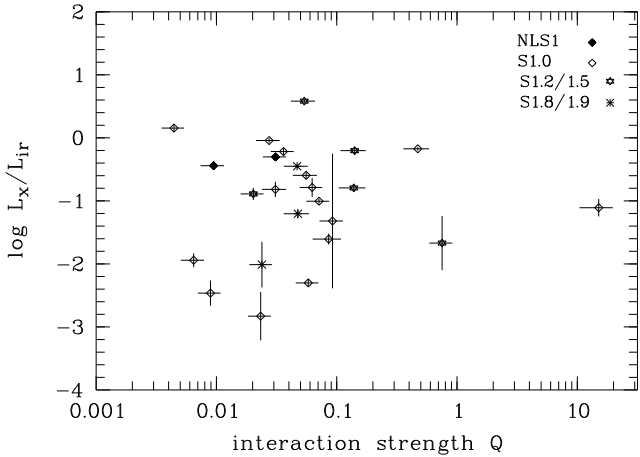,width=8.8cm,bbllx=15mm,bblly=15mm,bburx=200mm,bbury=275mm,angle=-90,clip=}}
\caption[]{Soft X-ray to far-infrared ($40-120 \rm \mu m$) luminosity ratio
$L_{\rm X}/L_{\rm fir}$ vs. dimensionless gravitational interaction strength $Q$
for all types of Seyfert 1.x galaxies. There is no correlation noticeable.}
\label{fig_Lx/Lir_Q}
\vspace{-0.5cm}
\end{center}
\end{figure}

In fig. \ref{fig_Lx/Lir_Q} we plot the ratio $L_{\rm X}/L_{\rm fir}$ vs. the interaction strength $Q$ for all Seyfert types. For Seyfert 1 galaxies
no correlation between the luminosity ratio and the interaction strength is found. 
If we can interprete the far-infrared luminosity as mainly caused by starburst activity, and the X-ray luminosity as mainly caused
by accretion processes, this indicates that starburst and AGN activity increase proportionally. 

The interpretations of the results are discussed in Sect. \ref{discussion}.


\subsection{Timing properties}

Below we discuss the long-term (time scales between half a year up to 8 years) X-ray and short-term 
(time scales of order up to a few days) variability of interacting and isolated Seyfert 1 and 2 galaxies.

\subsubsection{X-ray variable Seyfert~1 galaxies}
\label{varia_s1_sec}

In our sample $59\%$ of the X-ray detected Seyfert 1 
galaxies show significant X-ray variability during the ROSAT All-Sky Survey and ROSAT pointed observations.
The corresponding X-ray light curves are shown in Appendix \ref{tables_light}.

\begin{figure}[h]
\begin{center}
\centerline{\psfig{figure=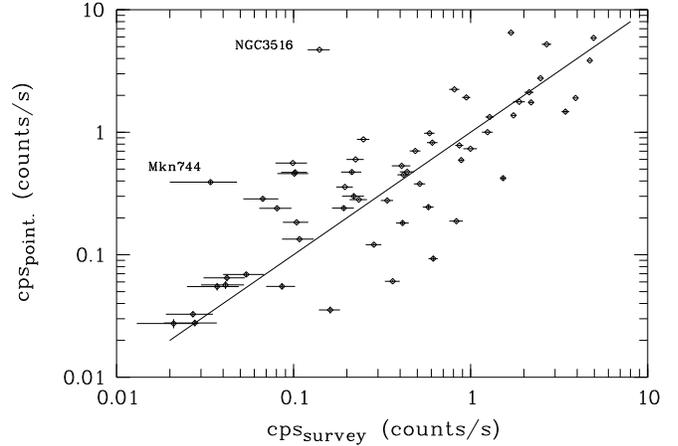,width=8.8cm,bbllx=15mm,bblly=15mm,bburx=200mm,bbury=280mm,angle=-90,clip=}}
\caption[]{Long-term variability of interacting and isolated Seyfert~1 galaxies. The most extreme 
factor of amplitude  variability of about 33  is found for NGC~3516.}
\label{fig_varias1}
\vspace{-0.5cm}
\end{center}
\end{figure}

In fig. \ref{fig_varias1} we compare the ROSAT All-Sky Survey count rate with the count rate measured
in ROSAT PSPC pointed observations. The most extreme factor of variability is found for NGC~3516 (a factor
of about 33 on a timescale of 718 days). 


\subsubsection{The X-ray light curves for  Seyfert 2 galaxies}
\label{varia_s2_sec}

\begin{figure}[h]
\begin{center}
\centerline{\psfig{figure=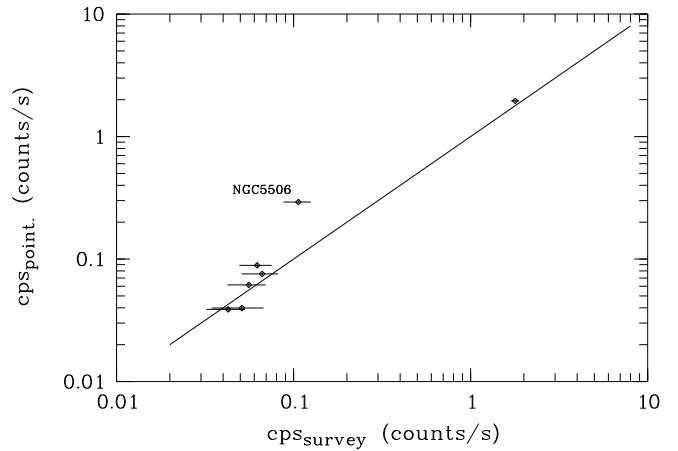,width=8.8cm,bbllx=15mm,bblly=15mm,bburx=200mm,bbury=280mm,angle=-90,clip=}}
\caption[]{The long-term variability of interacting and isolated Seyfert~2 galaxies.}
\label{fig_varias2}
\vspace{-0.8cm}
\end{center}
\end{figure}

For interacting and isolated Seyfert 2 galaxies 
no indication for significant X-ray variability on timescales above 0.5 years is found by comparing the
ROSAT All-Sky Survey and ROSAT PSPC pointed observations (Fig. \ref{fig_varias2}). The galaxy NGC 5506
is classified by Lipovetski et al. (1987) as Seyfert type 2. This source exhibits the largest factor of variability of
about 2.7 on a timescale of 375 days.

\begin{figure}[h]
\begin{center}
\vspace{0.0cm}
\vbox{
\centerline{\psfig{figure=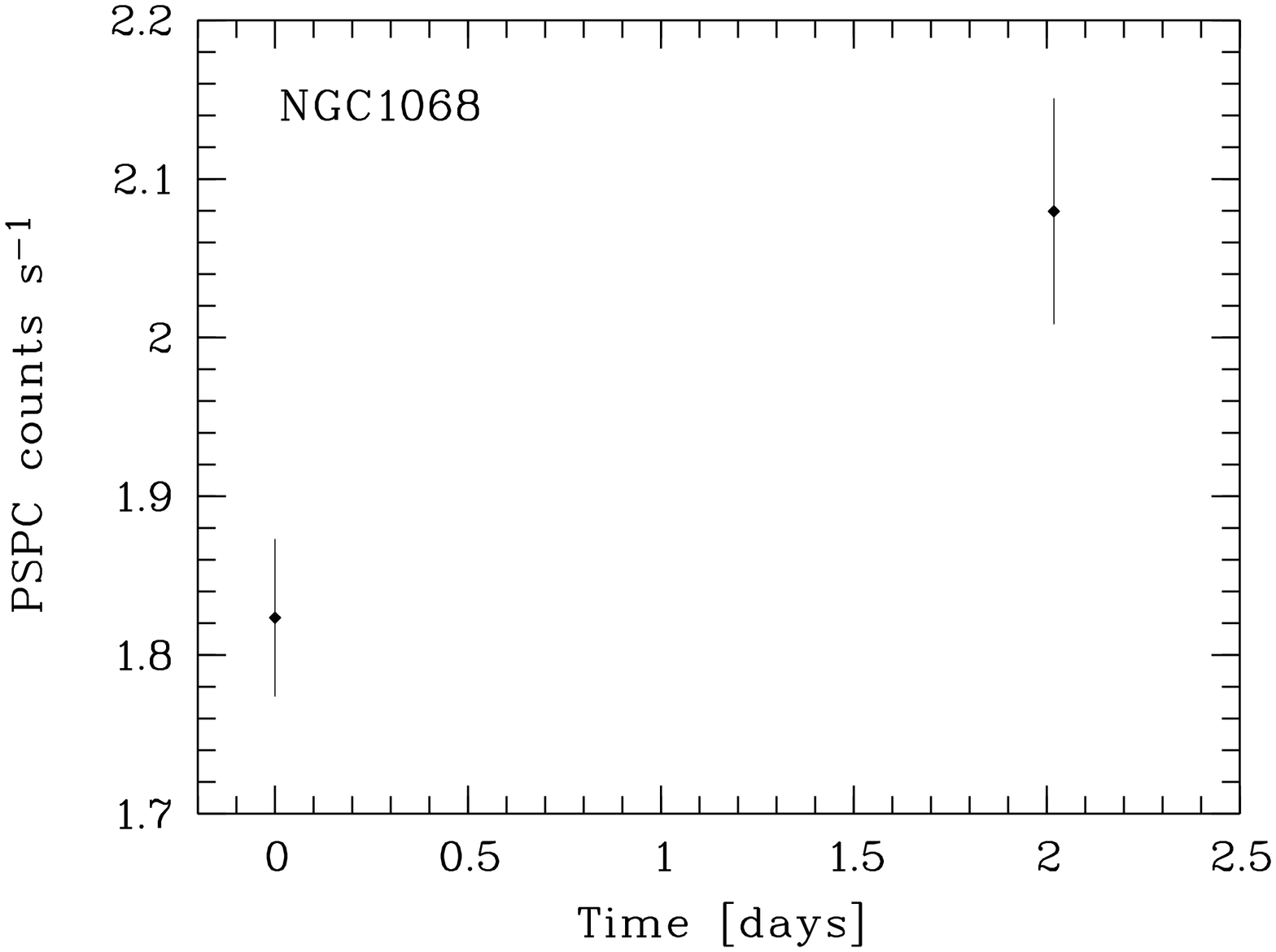,width=7.0cm,bbllx=15mm,bblly=10mm,bburx=180mm,bbury=240mm,angle=-90,clip=}}
\hspace{0.0cm}\\
\centerline{\psfig{figure=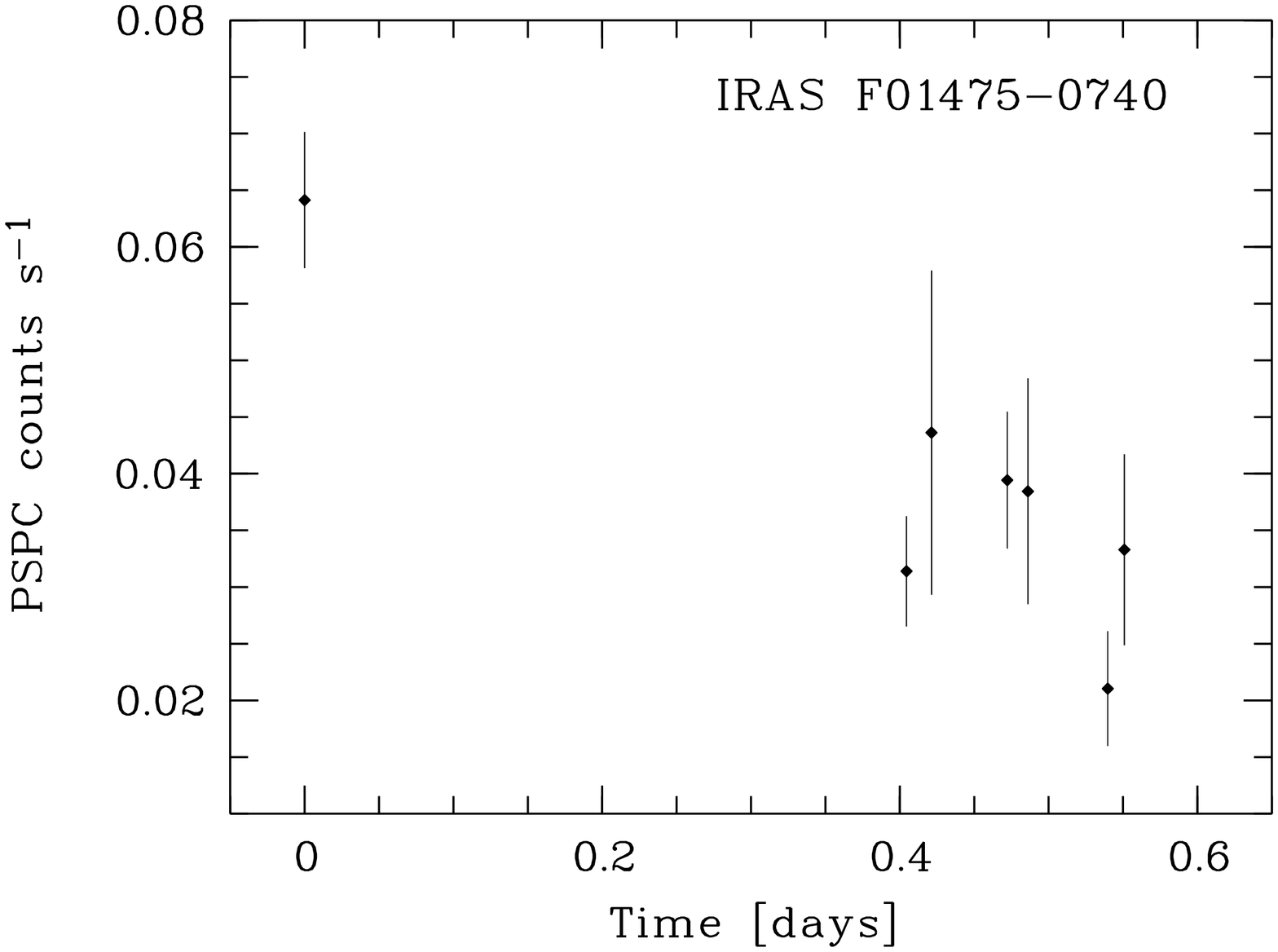,width=7.0cm,bbllx=15mm,bblly=10mm,bburx=180mm,bbury=240mm,angle=-90,clip=}}
\hspace{0.0cm}\\
\centerline{\psfig{figure=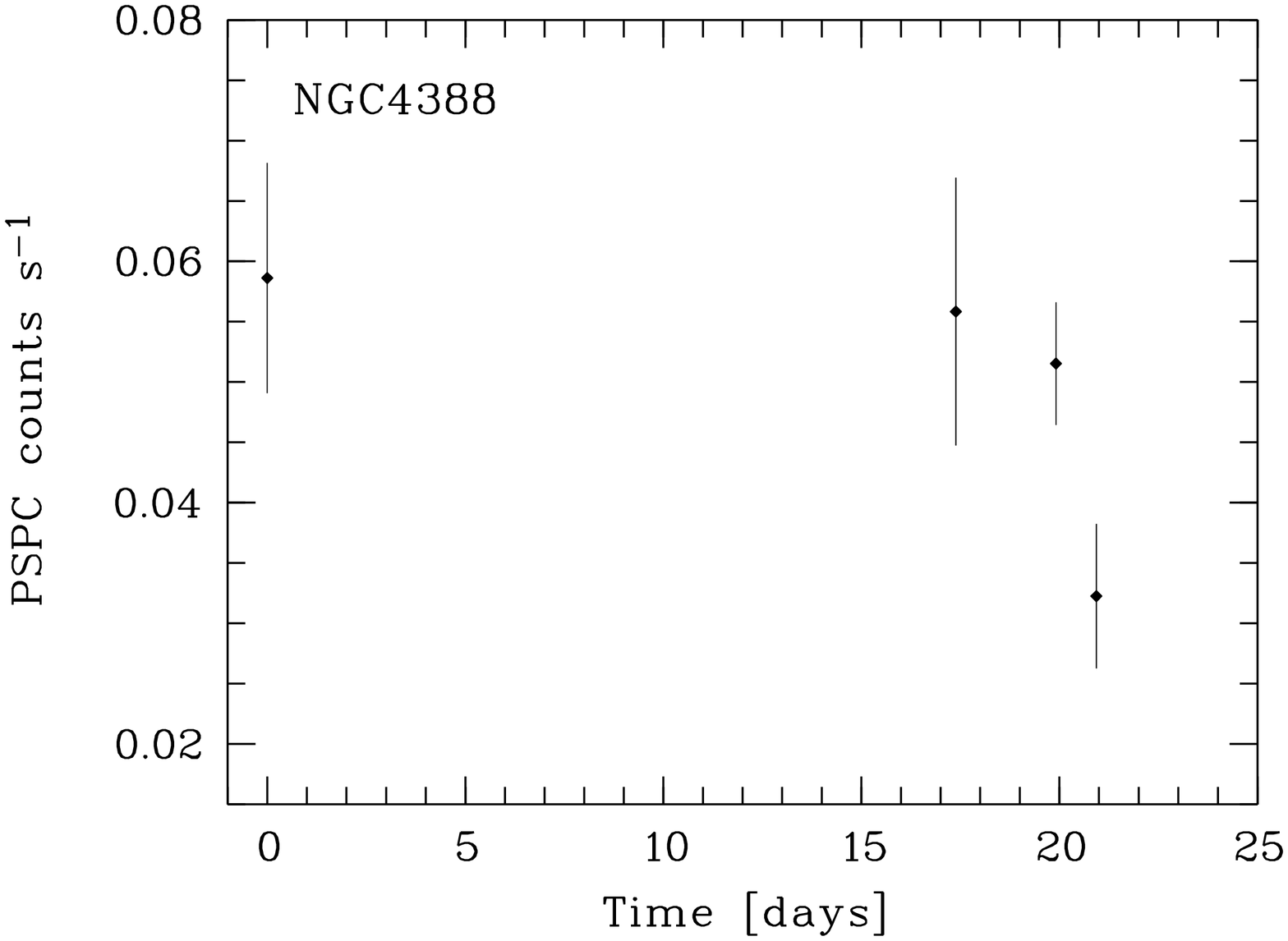,width=7.0cm,bbllx=15mm,bblly=10mm,bburx=180mm,bbury=240mm,angle=-90,clip=}}}
\caption[]{
PSPC pointed light curves of the Seyfert 2.0 galaxies NGC~1068 ({\em top}) with the probabilities of variability of 99.83$\%$
(3$\sigma$), IRAS~0147-0740 ({\em middle}) with 99.9989$\%$ (4$\sigma$) and NGC~4388 ({\em bottom}) with 97.896$\%$ (2$\sigma$).}
\label{fig_NGC1068}
\vspace{-0.8cm}
\end{center}
\end{figure}

However, for three out of the 36 Seyfert~2 galaxies, NGC~1068, NGC~4388 and
IRAS~F01475-0740, indications for  X-ray variability are found in ROSAT pointed observations.
In fig.\ref{fig_NGC1068} ({\em top}) the pointed observation light curve of the Seyfert 
2 galaxy NGC~1068 is shown.
An increase in count rate from 1.823 to 2.080 $\rm counts\;s^{-1}$, corresponding to a factor of 1.14 or
$\Delta\rm cps = 0.256$, within 2 days is detected. 
A constant model fit using the $\chi^2$ test can be rejected with a probability of $99.83\%$,
corresponding to 3$\sigma$. Indications for X-ray variability in NGC~1068 are also found  in
other \ros pointed observations (cf. fig. \ref{fig_NGC1068b}). 
The ROSAT PSPC light curve for the Seyfert 2 galaxy IRAS~01475-0748 is shown in 
fig. \ref{fig_NGC1068} ({\em middle}).
A decrease in the count rate from 0.064 to 0.021 $\rm counts\;s^{-1}$ within  12.9 hours is detected. 
This variability corresponds to a factor of 3 and to a change in the count rate of 
$\Delta\rm cps = 0.043$. A constant model fit gives 
a probability of 99.9989$\%$ (4$\sigma$).
In fig. \ref{fig_NGC1068} ({\em bottom}) the X-ray light curve of NGC~4388 is shown. 
The count rate decreases from 0.0586 to 0.0322 $\rm counts\ s^{-1}$  corresponding to a factor of variability of about 1.8
and a change in the count rate of 
$\Delta \rm cps = 0.026$ within 21 days. A constant model fit can be rejected with a probability
of 97.896$\%$, corresponding to 2$\sigma$.

Recently, Georgantopoulos \& Papadakis (2000) found evidence for spectral (and timing) variability for four Seyfert 2
galaxies in RXTE observations.


\section{Discussion and summary}
\label{discussion}

We have detected $92\%$ of interacting or isolated Seyfert~1 and $48\%$ of Seyfert~2 galaxies in the optically selected
sample of Rafanelli et al. (1995). The soft X-ray spectral and timing properties are presented in tables \ref{Xdats1.tab} \& \ref{Xdats2.tab} 
(note the different combination of Seyfert types; 102 S1.x and 36 S2).

For Seyfert 1 galaxies we have found a correlation between photon index $\Gamma$ and the soft X-ray
luminosity $L_{\rm X}$. A clear trend of a steepening of the X-ray spectrum with increasing X-ray luminosity is detected. 
High X-ray luminosity is most probably related to the accretion rate and/or the black hole mass.
Therefore, a possible explanation for this effect might be a shifted
and strengthened accretion disk spectrum in high-luminosity Seyferts.
We confirm that NLS1s have steeper X-ray continua than broad-line Seyfert 1 galaxies. 
Seyfert 2 galaxies show similar steep X-ray continua compared to NLS1s.
While the steep X-ray continua for NLS1s are expected to be related with high values for the Eddington luminosity in combination
with small black hole masses, the steep X-ray continua for Seyfert 2 galaxies
are probably due to the dominant X-ray line emission from the circumnuclear starburst.

Our data result in an increasing far-infrared and increasing X-ray luminosity with increasing interaction strength for Seyfert 1 galaxies.
Both strengthen the suggestion that galaxy interaction triggers an increased accretion rate and starburst rate.
NLS1s and Seyfert 2 galaxies show the highest values of far-infrared luminosity compared to Seyfert 1 galaxies. This fact points
to nuclear starburst activity taking place in NLS1 galaxies as suggested by Mathur (2000).

For Seyfert 2 galaxies, we found no significant correlations between the X-ray luminosity and photon index or interaction strength.
The main reason is the lack of penetration through the high column
densities in Seyfert 2 galaxies with soft X-rays. The ROSAT $N_{\rm H}$ values derived from the spectral fitting are therefore lower limits
to the true absorbing columns in Seyfert 2 galaxies. This is supported by comparing our results with those
from higher energy satellites. Bassani  et al. (1999) investigated the hard X-ray spectra of a large sample of
Seyfert 2 galaxies with {\sl Ginga}, {\sl ASCA} and {\sl BeppoSAX}. 
The authors found a large population
of strong-absorbed objects with column densities $N_{\rm H} \ge \rm 10^{23} cm^{-2}$, including many Compton-thick candidates. The
column densities obtained from the soft X-ray spectra show significantly lower values for  many sources. The reason for this
is the different origin of the hard and soft X-ray radiation. 
The column densities are likely able to determine the outer regions of the molecular torus at soft X-rays, 
because the radiation penetrates less absorbing material.
Only the hard X-ray radiation above a few keV is able to pass interior regions of the torus and leads to higher
column densities. Therefore, our results for Seyfert 2 galaxies are only a lower limit to the intrinsic column density. The fluxes and
luminosities of Seyfert 2 galaxies given in table \ref{Xdats2.tab} are considered as lower limits to the intrinsic values. 
Finally, the soft X-ray emission of Seyfert 2 galaxies is probably due to emission from the circumnuclear starburst
together with radiation from the nuclei, scattered and reflected by the molecular torus.
Moreover, we found no correlation between the far-infrared luminosity and interaction strength for the Seyfert 2 sample. \\
Higher sensitivity observations with {\sl XMM-Newton} and {\sl Chandra} are expected to confirm the short-time
variability of the three Seyfert 2 galaxies.

We have investigated the variability of Seyfert 1 and 2 galaxies on short and long timescales and found indications for
variability in three Seyfert 2.0 galaxies on short timescales (NGC~1068, IRAS~0147-0740, NGC~4388).
A possible explanation for this variability might be the presence of boreholes in the absorbing molecular torus around the
central black hole region. Significant X-ray variability during the \ros pointed and survey observations were detected for
58 percent of the Seyfert 1 galaxies.


\begin{acknowledgements}

We thank Prof. Joachim Tr\"umper for comments and A. Vogler for help in producing the images with overlayed X-ray contours.
We are grateful to Dr. W. Voges for his help in the catalogue preparation.
The \ros project is supported by the Bundesministerium f\"ur Bildung und Forschung (BMBF/DLR) and by the
Max-Planck Society (MPG).\\
This paper can be retrieved via WWW from our pre-print server: http://www.xray.mpe.mpg.de/$\sim$pfefferk/

\end{acknowledgements}



\begin{appendix}

\section{Notes on individual sources}
\label{notes}

\begin{itemize}
      \item[$\bullet$]{\bf NGC~2992}: In the case of NGC~2992 and its companion we detect two spatially separated X-ray
components. In table \ref{Xdats2.tab} of Appendix \ref{tables_light} we give the sum of the count rates, fluxes and luminosities for
this system. For the completion of the database we add here the individual values of the source and the companion. The spectrum of
NGC2992 shows a high absorption and a low photon index with an integrated flux
of $2.815\cdot10^{-12}\;\rm erg\;cm^{-2}\;s^{-1}$
corresponding to a luminosity of $2.616\cdot10^{41}\;\rm erg\;s^{-1}$. The spectrum of the companion results in a flux of
$1.061\cdot10^{-12}\;\rm erg\;cm^{-2}\;s^{-1}$ corresponding to a luminosity of
$9.987\cdot10^{40}\;\rm erg\;s^{-1}$.
The spectral fit parameters for the companion are $N_{\rm
H_{fit}}=0.854\cdot10^{21}\rm cm^{-2}$ and $\Gamma=-2.72$.\\
NGC~2992 is classified by Lipovetski et al. (1987) as a Seyfert 1.9 galaxy and we found a variability of this object over short
and long timescales (see Sect. \ref{varia_s2_sec}).
      \item[$\bullet$]{\bf NGC~5506}: The soft X-ray spectrum of the Seyfert galaxy NGC~5506 is highly absorbed and no reliable
spectral fit parameters can be obtained. Thus, we have used for
$N_{\rm H}=1.0\cdot10^{21}\rm cm^{-2}$ as a lower limit in
table \ref{Xdats2.tab}.
      \item[$\bullet$]{\bf NGC~5953}: This interacting system (Rafanelli, Osterbrock, Pogge; 1990) is detected in the survey
and pointed observations. The X-ray emission of this system in the pointed observation is mainly caused by the Seyfert 2 galaxy
NGC~5953. In the survey observation the emission is centered on the companion of NGC~5953 and no emission from the Seyfert 2 galaxy is detected.
Due to the PSPC pointing accuracy we are not able to decide whether NGC~5953 is a transient source in the soft X-ray band.
      \item[$\bullet$]{\bf Mkn~684}: For Mkn~684 we assume the classification as a NLS1 galaxy by Osterbrock \& Pogge (1985) and Grupe (1996).
      \item[$\bullet$]{\bf NGC~7319}, a member of the Stefan's Quintet group, is detected with the PSPC (pointed and survey)
and HRI detector by {\sl ROSAT}. The PSPC survey and pointed observations showed unresolved X-ray emission from the group and the 
intergalactic gas (Pietsch W. \& Trinchieri G., 1997). The HRI pointed observation as used to determine the flux of the Seyfert 2
galaxy.
      \item[$\bullet$]{\bf NGC~3031}: For NGC~3031 (M81) we used the distance of 3.63 Mpc given in Freedman et al. (1994) to compute
the luminosity of this object ($z < 0$).
   \end{itemize}

\section{Light curves of NGC~1068}
\label{NGC1068_light}

\begin{figure}[h]
\begin{center}
\vbox{
\centerline{\psfig{figure=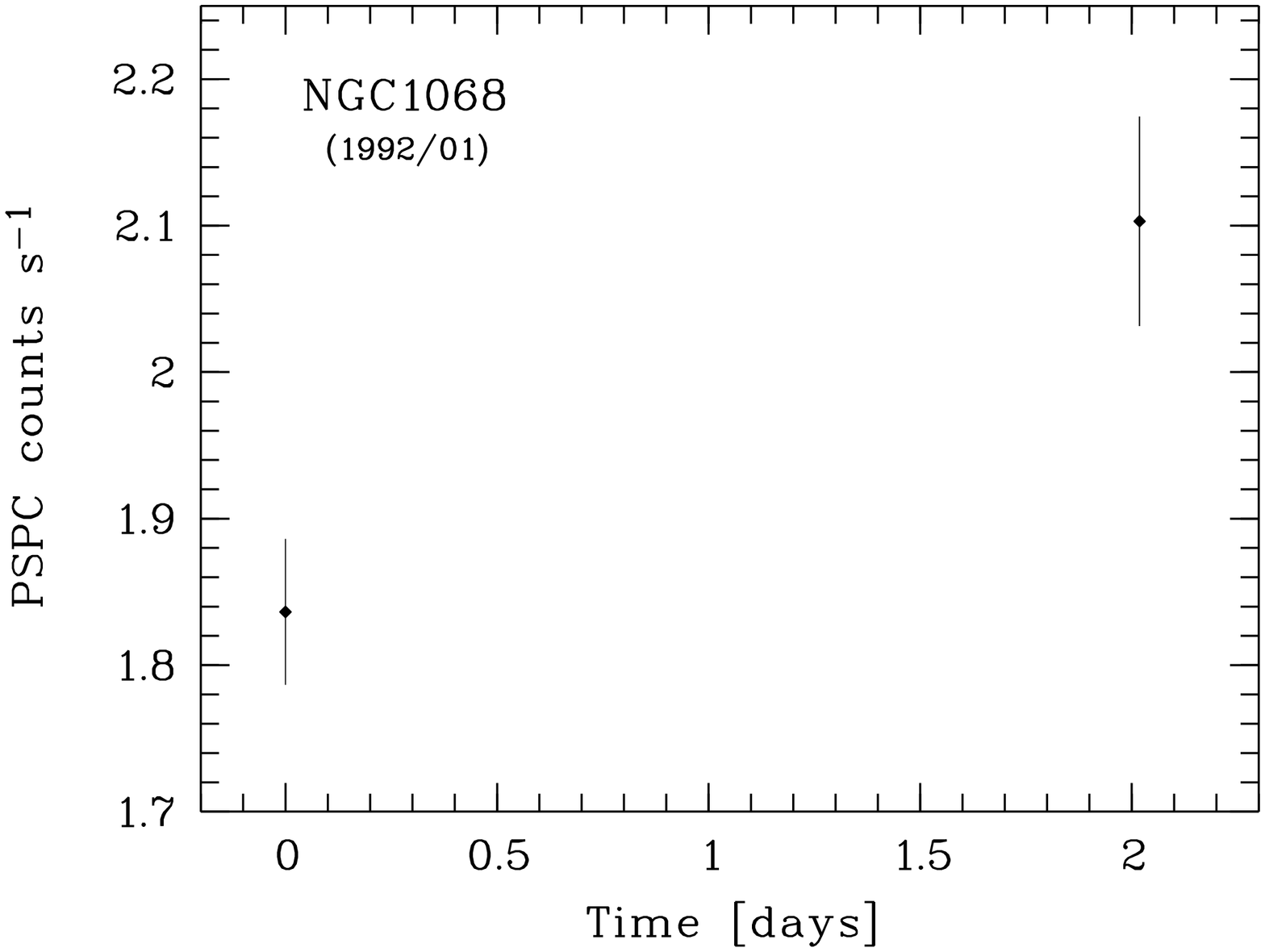,height=4.0cm,width=7.5cm,bbllx=15mm,bblly=10mm,bburx=180mm,bbury=240mm,angle=-90,clip=}}
\centerline{\psfig{figure=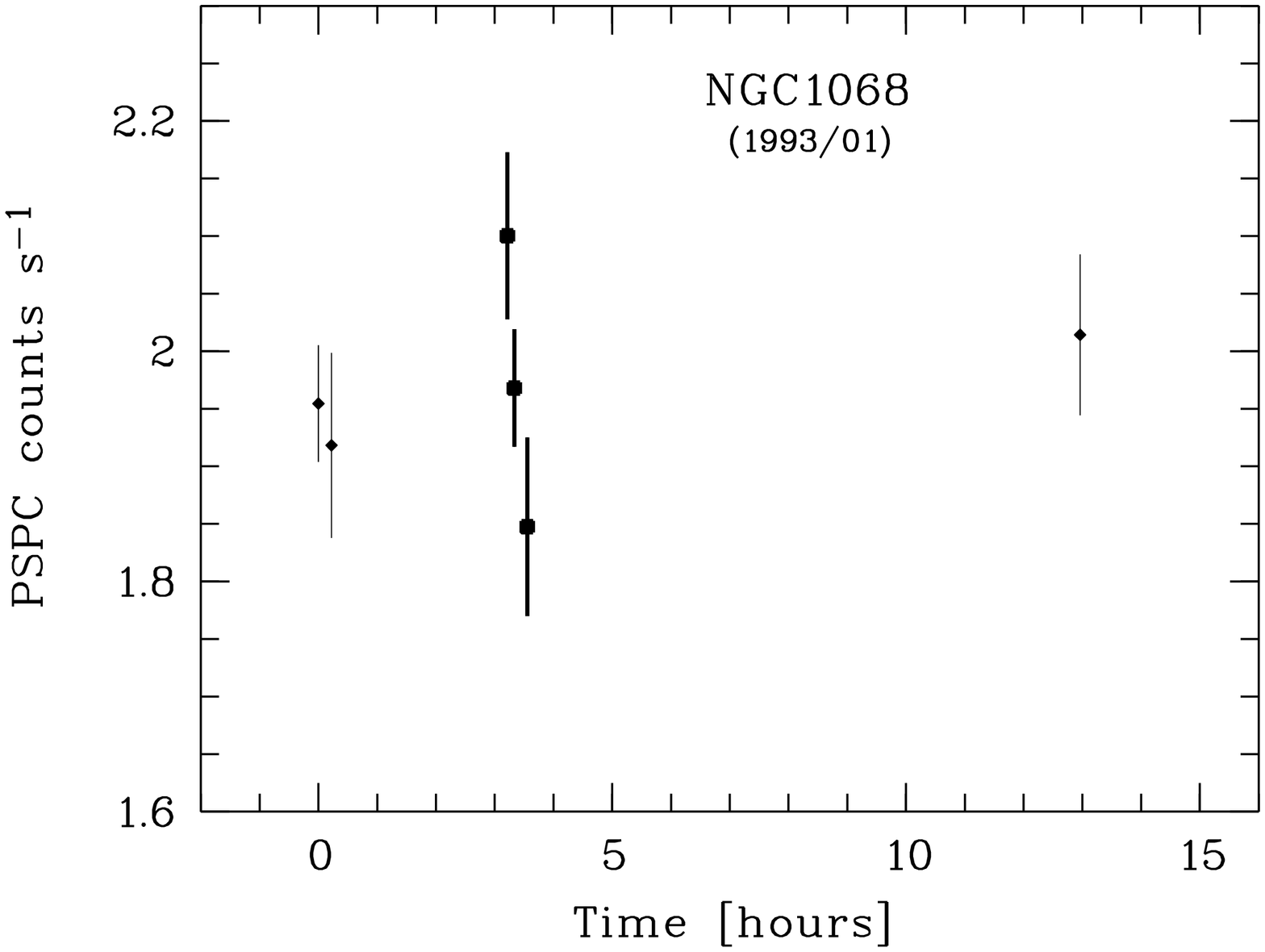,height=4.0cm,width=7.5cm,bbllx=15mm,bblly=10mm,bburx=180mm,bbury=240mm,angle=-90,clip=}}
\centerline{\psfig{figure=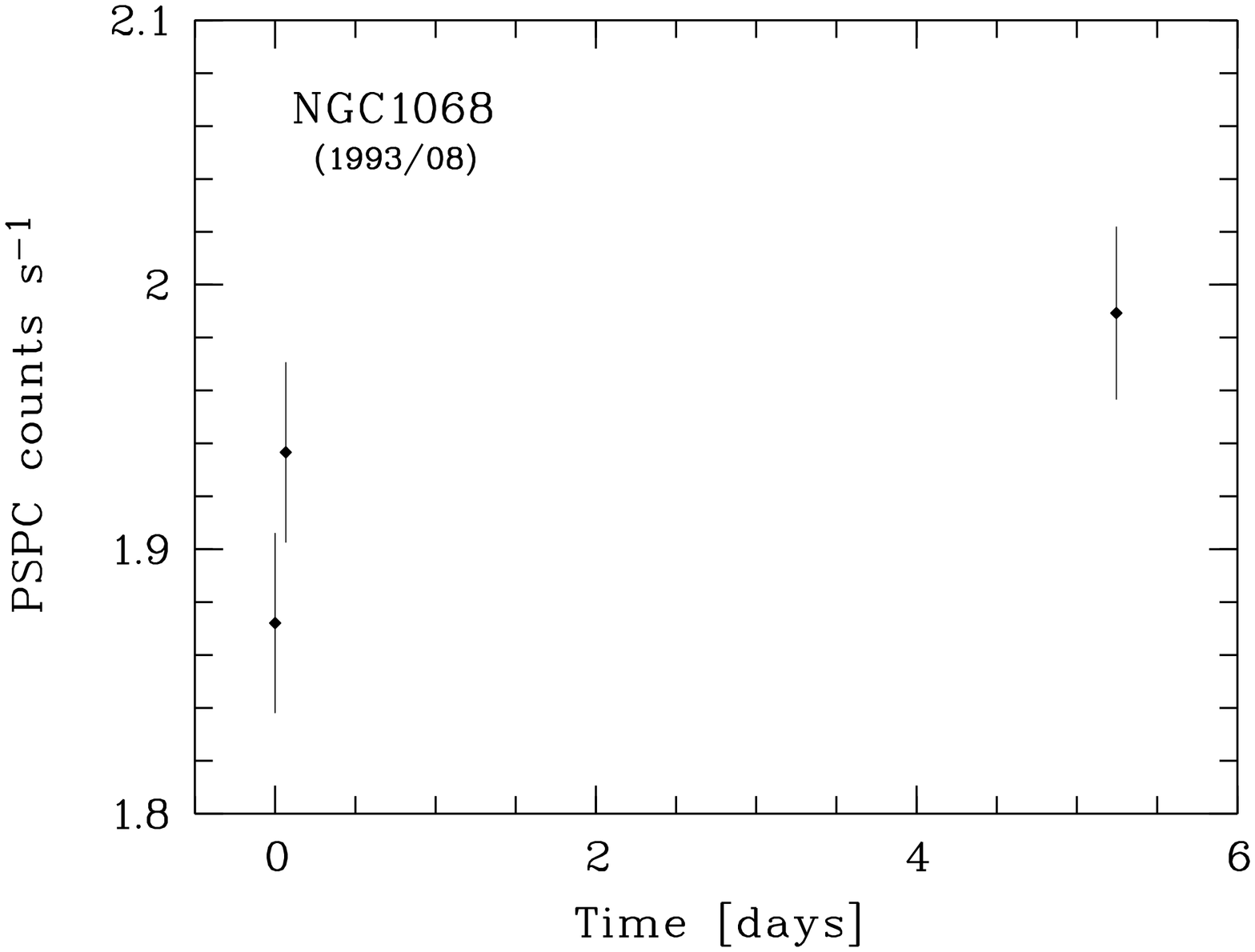,height=4.0cm,width=7.5cm,bbllx=15mm,bblly=10mm,bburx=180mm,bbury=240mm,angle=-90,clip=}}
\centerline{\psfig{figure=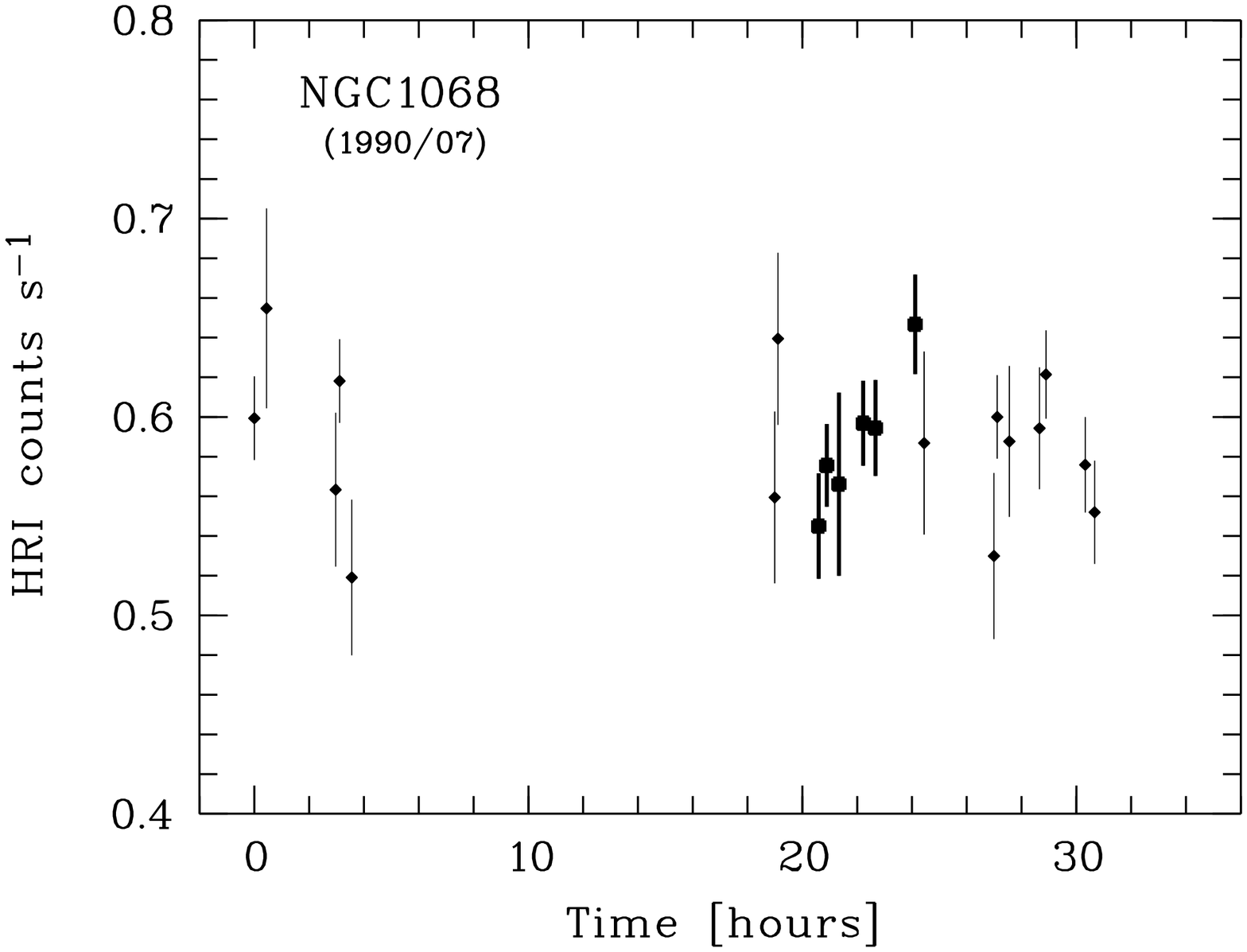,height=4.0cm,width=7.5cm,bbllx=15mm,bblly=10mm,bburx=180mm,bbury=240mm,angle=-90,clip=}}}
\caption[]{Pointing light curves of the Seyfert 2.0 galaxies NGC~1068 (from top to bottom): {\em\underline{first panel}}:
this PSPC light curve of NGC1068 from 1992/01 was described in Sect. \ref{varia_s2_sec} already. {\em\underline{second panel}}:
The bold data points in the light curve of the PSPC observation from 1993/01 indicate a variability with a probability of
$94.24\%$ ($\sim2\sigma$). The count rate decrease from 2.10 to 1.85 $\rm counts\;s^{-1}$ corresponds to a factor of 1.14 or
$\Delta\rm cps=0.253$ within 1 hour. {\em\underline{third panel}}: The third PSPC pointing light curve of NGC~1068 shows a
probability of variability of $95.47\%$ ($2\sigma$) with an increase of
$\Delta\rm cps=0.117$ $\rm counts\;s^{-1}$.
{\em\underline{fourth panel}}: A HRI pointed light curve of this object indicates also a variability, with a probability
of $88,82\%$ for the bold data points. The count rate increase from 0.545 to
0.647 $\rm counts\;s^{-1}$ within 3.5 hours correlates
with a variability factor of 1.19 or  $\Delta\rm cps=0.102$.}
\label{fig_NGC1068b}
\vspace{-0.5cm}
\end{center}
\end{figure}

In Fig. \ref{fig_NGC1068b} we give the remaining PSPC and HRI light curves of the Seyfert 2 galaxy NGC~1068. 


\section{The Catalogue - data tables and light curves}
\label{tables_light}

\begin{figure*}[h]
\begin{center}
\centerline{\psfig{figure=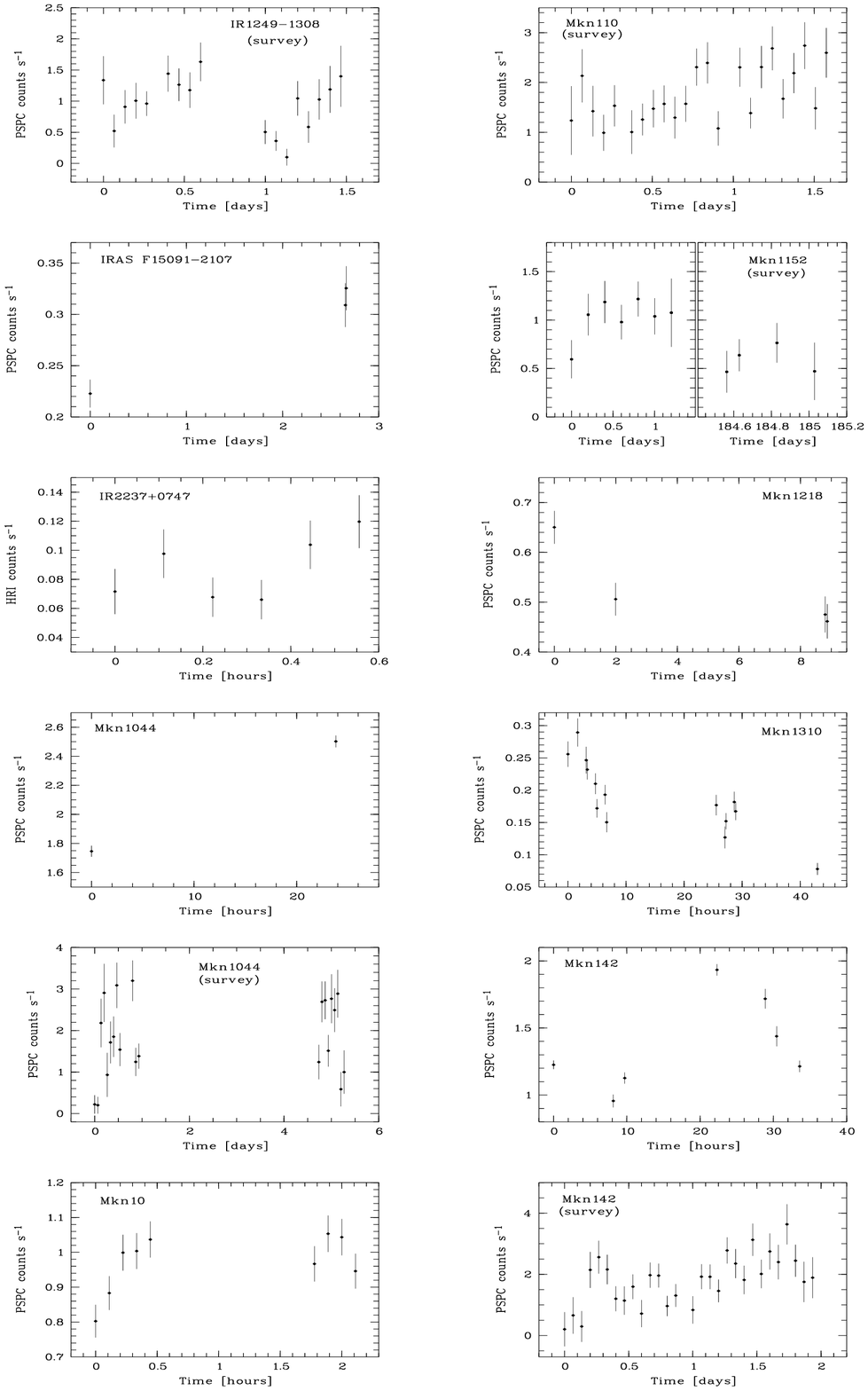,width=16.0cm,bbllx=25mm,bblly=15mm,bburx=190mm,bbury=275mm,clip=}}
\caption[]{light curves of variable Seyfert 1 galaxies}
\label{fig_light1}
\vspace{-1.0cm}
\end{center}
\end{figure*}

\begin{figure*}[h]
\begin{center}
\centerline{\psfig{figure=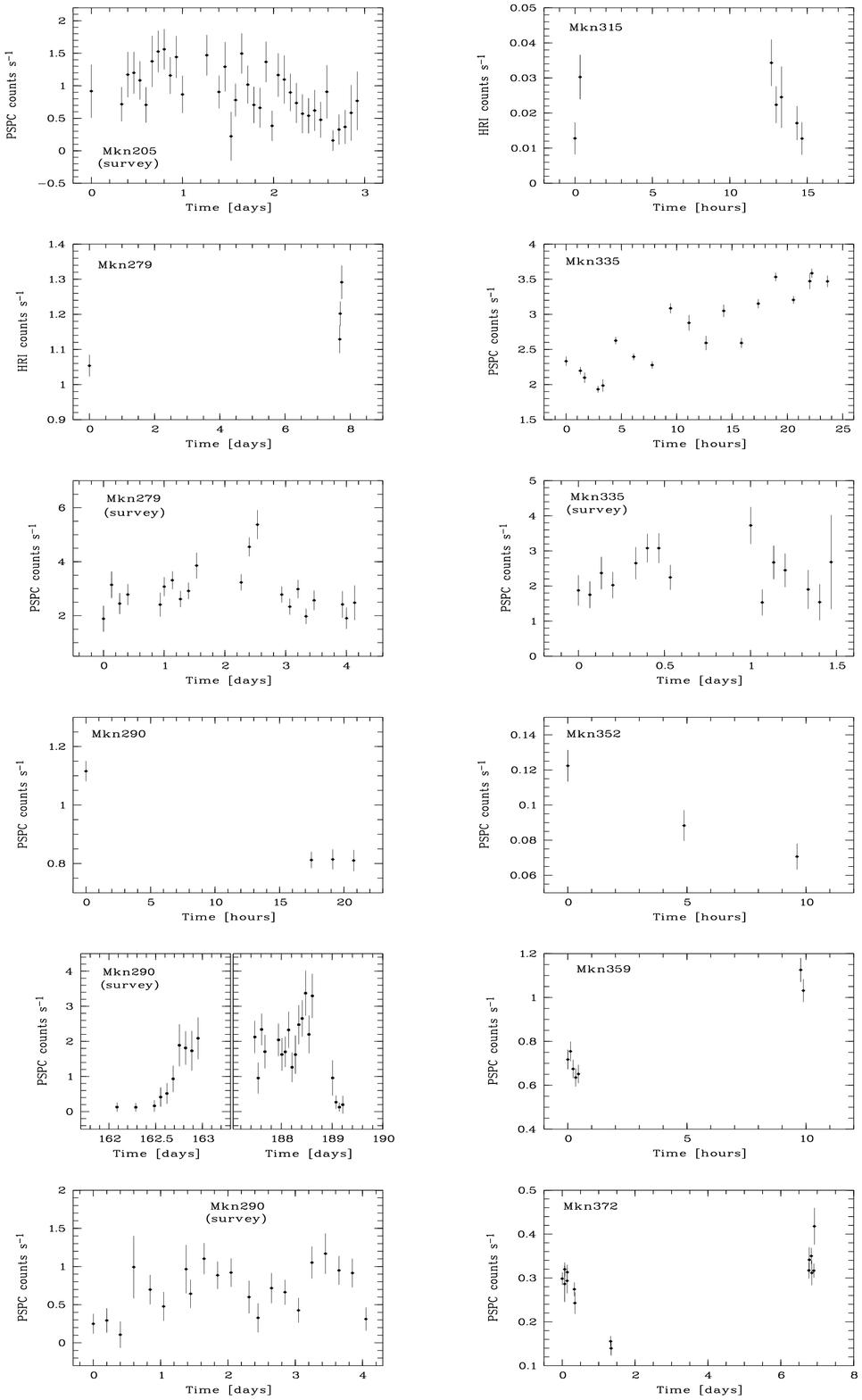,width=16.0cm,bbllx=25mm,bblly=15mm,bburx=190mm,bbury=275mm,clip=}}
\caption[]{light curves of variable Seyfert 1 galaxies}
\label{fig_light2}
\vspace{-1.0cm}
\end{center}
\end{figure*}

\begin{figure*}[h]
\begin{center}
\centerline{\psfig{figure=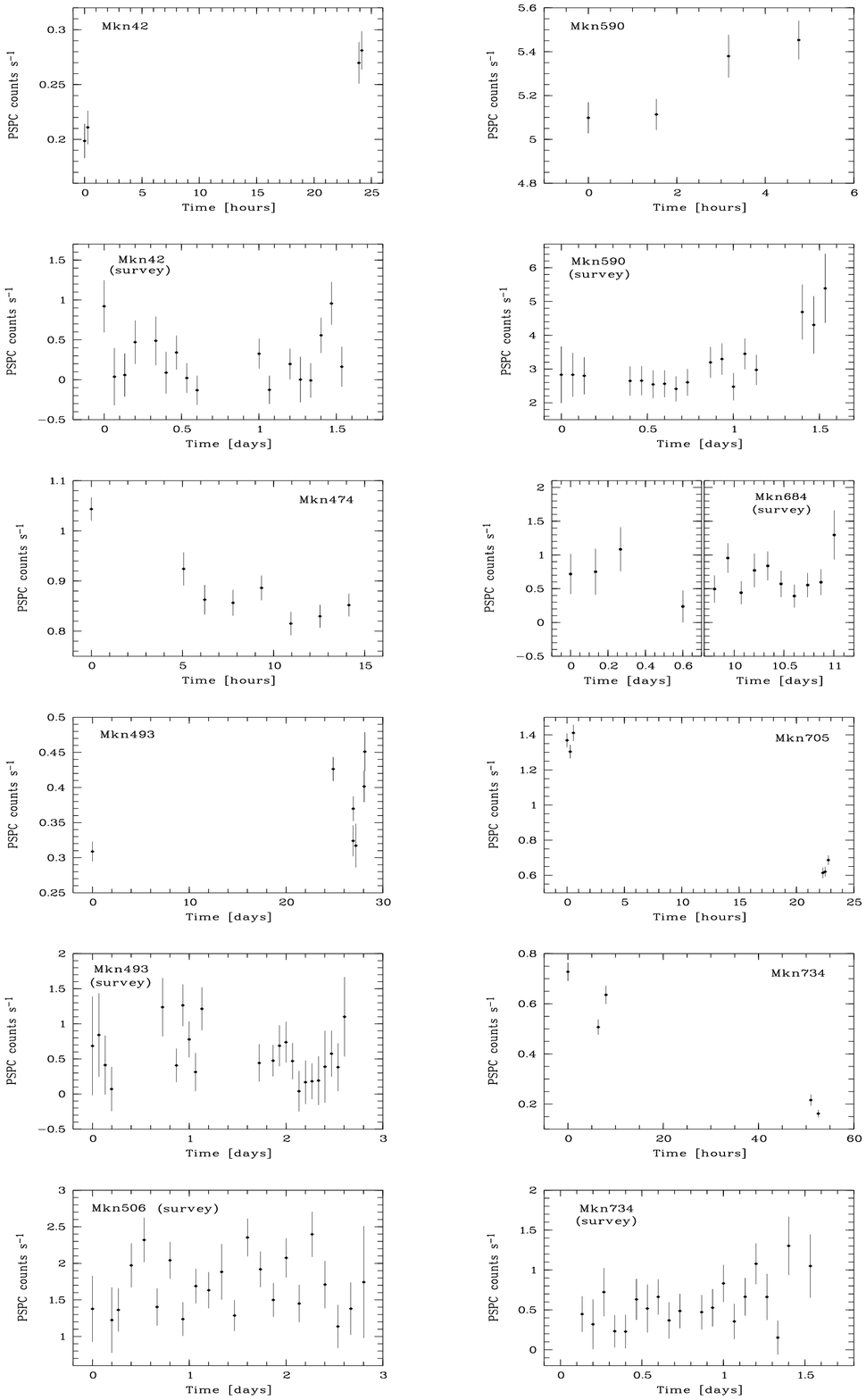,width=16.0cm,bbllx=25mm,bblly=15mm,bburx=190mm,bbury=275mm,clip=}}
\caption[]{light curves of variable Seyfert 1 galaxies}
\label{fig_light3}
\vspace{-1.0cm}
\end{center}
\end{figure*}

\begin{figure*}[h]
\begin{center}
\centerline{\psfig{figure=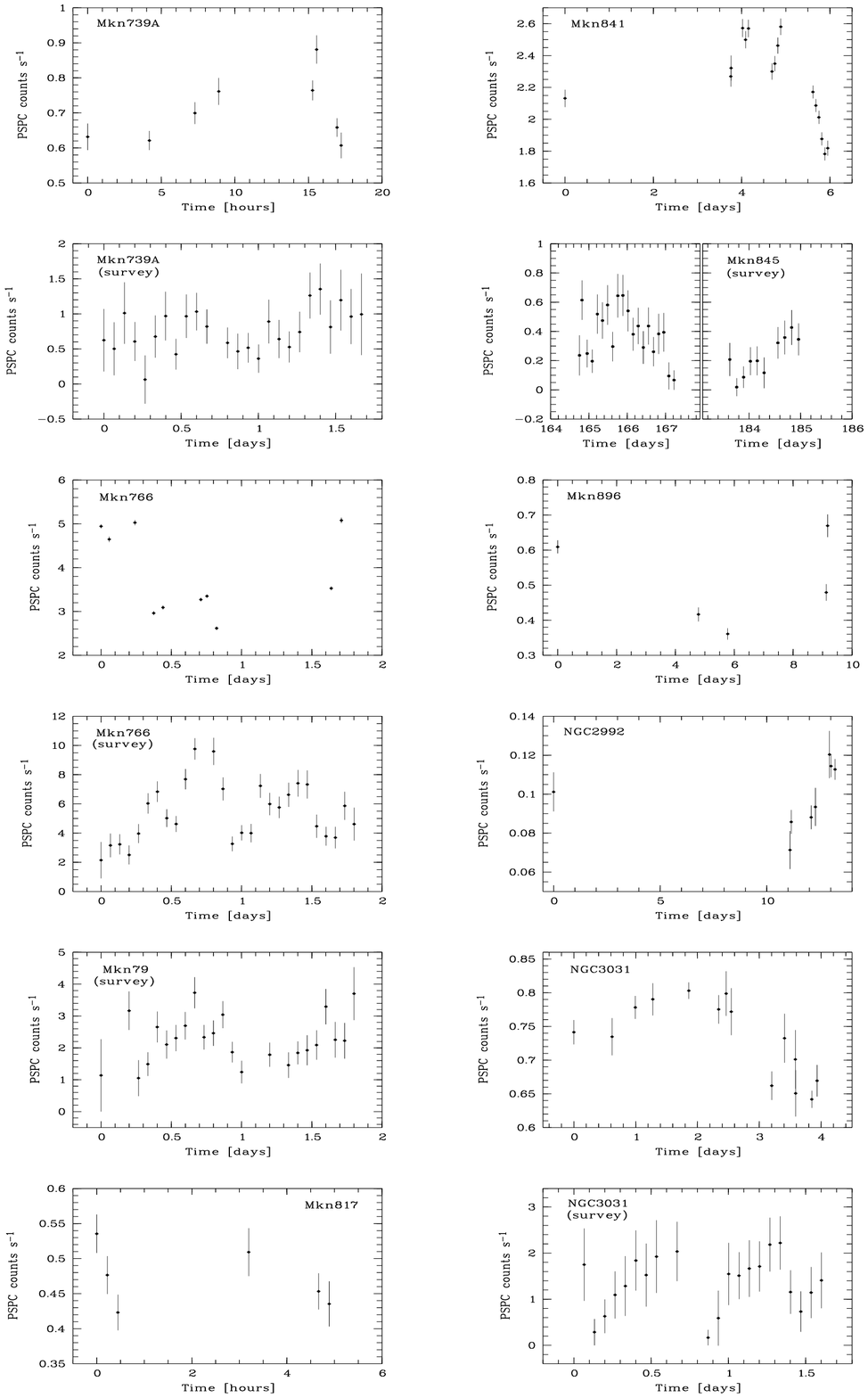,width=16.0cm,bbllx=25mm,bblly=15mm,bburx=190mm,bbury=275mm,clip=}}
\caption[]{light curves of variable Seyfert 1 galaxies}
\label{fig_light4}
\vspace{-1.0cm}
\end{center}
\end{figure*}

\begin{figure*}[h]
\begin{center}
\centerline{\psfig{figure=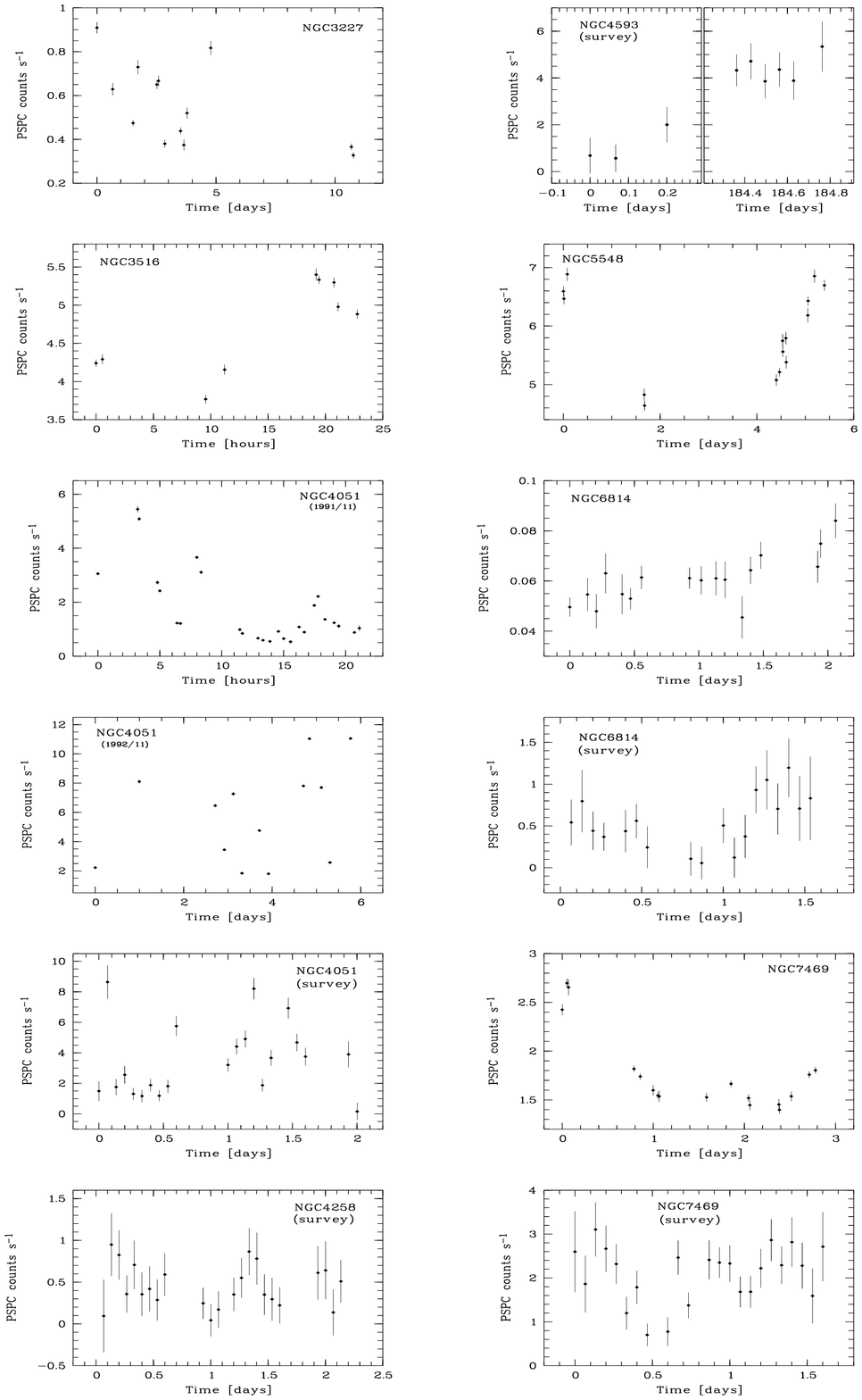,width=16.0cm,bbllx=25mm,bblly=15mm,bburx=190mm,bbury=275mm,clip=}}
\caption[]{light curves of variable Seyfert 1 galaxies}
\label{fig_light5}
\vspace{-1.0cm}
\end{center}
\end{figure*}

In this section we show the results of our studies of the X-ray properties of the Seyfert 1 and 2 galaxies in tables \ref{Xdats1.tab} \&
\ref{Xdats2.tab}, while in tables \ref{rafdats1.tab} \& \ref{rafdats2.tab}, general and optical properties are listed.
The light curves of variable Seyfert 1 galaxies are also summarized in this appendix in Figs. \ref{fig_light1} to \ref{fig_light5}.
To distinguish the survey from the pointing light curves we have labeled the survey curves. The pointed light curves are
distinguished by the y-axis in PSPC or HRI based data.

\clearpage


\begin{table*}
\caption{\bf\underline{Seyfert1 galaxies:} \rm\small{A part of the Rafanelli sample of interacting or isolated
Seyfert 1 galaxies detected by {\em ROSAT}. The table contains optical and generical properties as well
as some results from the investigations of the Rafanelli group (Rafanelli et al., 1995). Col(2) - object
name, col(3) - ROSAT name, col(4) - redshift, col(5\&6) - diameter of Seyfert galaxy and companion, col(7) - distance between
the components, col(8) - interaction strength, col(9) - apparent visual magnitude, col(10\&11) - quality
of X-ray identification (1: high, 2: lower degree of reliability) and col(12) -
Seyfert type. The upper index $\#$ in col(12) marks objects, which 
are classified as NLS1 galaxy (D.E. Osterbrock \& R.W. Pogge (1985); Th. Boller, W.N.Brandt \& H.Fink 1996; D.Grupe 1996).}}
\scriptsize
\begin{center}
\begin{tabular}{rllccccccccc} \hline\hline

Nr. & \bf name & \bf ROSAT name & $z$ & $D_{\rm p}$ & $D_{\rm c}$ & $S$ & $Q$ & $V$ &\multicolumn{2}{c}{id} & Sy \\

& & & & mm & mm & mm & & mag & point.& sur. & type \\ \hline

  1 & Mkn334 & 1RXS J000308.6$+$215728  & 0.022 & 3.7 & - & - & - & 14.62 & - & 1 & 1.8 \\ 
  2 & Mkn335 & 1RXP J000619.2$+$201224  & 0.025 & 1.2 & - & - & - & 13.85 & 1 & 1 & 1.0 \\
  3 & Mkn1146 & 1RXS J004719.4$+$144215 & 0.039 & 2.6 & - & - & - & 15.28 & - & 1 & 1.0 \\
  4 & Mkn352   & 1RXP J005953.3$+$314947 & 0.015 & 1.4 & - & - & - & 14.66 & 1 & 1 & 1.0 \\
  5 & Mkn1152  & 1RXP J011350.1$-$145034 & 0.052 & 2.2 & 1.2 & 1.3 & 1.9524$\pm$0.7060 & 15.00 & 1 & 1 & 1.5 \\
  6 & Mkn993 & 1RXS J012531.4$+$320800 & 0.017 & 7.5 & - & - & - & 13.96 & - & 1 & 1.9 \\
  7 & Mkn975  & 1RXS J011350.4$+$131534 & 0.050 & 2.0 & - & - & - & 14.50 & - & 1 & 1.5 \\
  8 & Mkn358   & 1RXS J012634.2$+$313659 & 0.042 & 2.8 & - & - & - & 15.23 & - & 1 & 1.0 \\
  9 & Mkn359   & 1RXP J012732.8$+$191051 & 0.017 & 2.6 & - & - & - & 14.21 & 1 & 1 & $1.5^{\#}$ \\
 10 & Mkn1018 & 1RXP J020616.0$-$001732 & 0.043 & 2.6 & - & - & - & 14.65 & 1 & 1 & 1.9 \\
 11 & Mkn590  &  1RXP J021433.7$-$004604 & 0.027 & 4.1 & 0.8 & 4.8 & 0.0537$\pm$0.0122 & 13.55 & 1 & 1 & 1.5 \\
 12 & Mkn1040  & 1RXS J022814.6$+$311838  & 0.016 & 9.8 & 1.8 & 1.7 & 15.0800$\pm$4.6514 & 14.74 & - & 1 & 1.0 \\
 13 & Mkn1044   & 1RXP J023005.8$-$085940 & 0.016 & 1.7 & - & - & - & 14.67 & 1 & 1 & $1.0^{\#}$ \\
 14 & NGC985  & 1RXP J023437.9$-$084709 & 0.043 & 3.2 & - & - & - & 14.95 & 1 & 1 & 1.0 \\
 15 & Mkn595   & 1RXS J024135.2$+$071117 & 0.028 & 1.7 & 1.0 & 2.0 & 0.2771$\pm$0.0790 & 14.40 & - & 1 & 1.0 \\
 16 & Mkn372 & 1RXP J024920.6$+$191822 & 0.031 & 1.7 & - & - & - & 14.81 & 1 & 1 & 1.8 \\
 17 & Mkn609 & 1RXP J032525.5$-$060827 & 0.032 & 1.1 & - & - & - & 14.12 & 1 & 1 & 1.8 \\
 18 & Mkn618  & 1RXS J043622.3$-$102226 & 0.035 & 2.6 & 0.7 & 4.3 & 0.0309$\pm$0.0071 & 14.65 & 1 & 1 & 1.0 \\
 19 & Mkn6   & 1RXS J065209.8$+$742537 & 0.019 & 1.8 & - & - & - & 14.02 & - & 1 & 1.5 \\
 20 & Mkn374   & 1RXS J065938.5$+$541136 & 0.044 & 1.6 & - & - & - & 14.38 & - & 2 & 1.2 \\
 21 & Mkn9   & 1RXS J073657.0$+$584610  & 0.039 & 1.7 & - & - & - & 14.68 & - & 1 & 1.0 \\
 22 & Mkn79   & 1RXP J074233.5$+$494838 & 0.022 & 3.3 & - & - & - & 14.78 & 1 & 1 & 1.2 \\
 23 & Mkn10   & 1RXP J074728.6$+$605552 & 0.030 & 5.2 & 2.1 & 10.0 & 0.0361$\pm$0.0078 & 14.70 & 1 & 1 & 1.0 \\
 24 & Mkn382    & 1RXS J075526.1$+$391111 & 0.034 & 2.1 & - & - & - & 15.50 & 1 & 1 & 1.0 \\
 25 & Mkn1218 & 1RXP J083810.8$+$245336 & 0.028 & 2.6 & 1.3 & 5.1 & 0.0468$\pm$0.0105 & 14.12 & 1 & 1 & 1.8 \\
 26 & NGC2639   & 1RXP J084338.0$+$501207 & 0.011 & 6.5 & - & - & - & 11.39 & 1 & - & 1.0 \\
 27 & NGC2782   &  1RXH J091405.1$+$400651 & 0.008 & 8.5 & 1.5 & 12.5 & 0.0233$\pm$0.0050 & 13.45 & 1 & - & 1.0 \\
 28 & Mkn704    & 1RXS J091826.2$+$161825 & 0.029 & 2.5 & 1.1 &  5.5 & 0.0274$\pm$0.0061 & 14.12 & - & 1 & 1.0 \\
 29 & Mkn110   & 1RXP J092512.8$+$521712 & 0.036 & 1.2 & - & - & - & 15.37 & 1 & 1 & 1.0 \\
 30 & Mkn705   & 1RXP J092603.0$+$124359  & 0.028 & 2.1 & - & - & - & 14.55 & 1 & 1 & $1.0^{\#}$ \\
 31 & NGC2992 & 1RXP J094541.6$-$141938 & 0.007 & 5.6 & 2.7 & 13.5 & 0.0239$\pm$0.0051 & 13.78 & 1 & 1 & 1.9 \\
 32 & Mkn124    & 1RXS J094841.6$+$502926 & 0.056 & 1.0 & - & - & - & 15.33 & - & 1 & 1.0 \\
 33 & Mkn1239   & 1RXP J095219.0$-$013631 & 0.019 & 1.3 & - & - & - & 14.33 & 1 & 1 & $1.5^{\#}$ \\
 34 & NGC3031  & 1RXP J095532.8$+$690354  & 0.000 & 6.9 & - & - & - & 11.72 & 1 & 1 & 1.5 \\
 35 & NGC3080   & 1RXP J095956.3$+$130242 & 0.035 & 2.2 & - & - & - & 15.01 & 1 & 1 & 1.0 \\
 36 & NGC3185  & 1RXP J101737.8$+$214124 & 0.004 & 0.7 & - & - & - & 12.73 & 1 & - & 1.0 \\
 37 & Mkn141  & 1RXS J101912.1$+$635802 & 0.039 & 1.5 & 1.4 &  2.8 & 0.1386$\pm$0.0349 & 15.27 & 1 & 1 & 1.2 \\
 38 & NGC3227  & 1RXP J102330.2$+$195150  & 0.003 & 12.5 & 8.0 & 11.0 & 0.7513$\pm$0.1615 & 11.29 & 1 & 1 & 1.0 \\
 39 & Mkn142   & 1RXP J102532.5$+$514045 & 0.045 & 1.0 & - & - & - & 15.50 & 1 & 1 & $1.0^{\#}$ \\
 40 & Mkn634    & 1RXS J105801.2$+$202937 & 0.066 & 1.8 & - & - & - & 15.49 & - & 2 & 1.0 \\
 41 & NGC3516  &  1RXP J110649.0$+$723406 & 0.009 & 7.0 & 1.7 & 21.0 & 0.0044$\pm$0.0009 & 12.23 & 1 & 1 & 1.0 \\
 42 & Mkn732   & 1RXS J111349.5$+$093518  & 0.030 & 2.6 & - & - & - & 14.00 & - & 1 & 1.0 \\
 43 & Mkn734  & 1RXP J112146.9$+$114418   & 0.049 & 1.1 & - & - & - & 14.71 & 1 & 1 & $1.0^{\#}$ \\
 44 & Mkn40    & 1RXP J112536.6$+$542303  & 0.020 & 1.3 & 0.7 &  1.5 & 0.2572$\pm$0.0851 & 15.39 & 1 & 2 & 1.0 \\
 45 & Mkn739A  &  1RXP J113629.2$+$213543 & 0.030 & 1.8 & - & - & - & 14.02 & 1 & 1 & 1.0 \\
 46 & Mkn744 & 1RXP J113942.8$+$315439 & 0.010 & 4.8 & 7.5 & 6.8 & 0.6870$\pm$0.1507 & 13.74 & 1 & 1 & 1.8 \\
 47 & NGC3884  & 1RXS J114611.5$+$202355  & 0.023 & 6.7 & - & - & - & 12.40 & - & 1 & 1.0 \\
 48 & Mkn42    &  1RXP J115341.7$+$461254 & 0.024 & 2.1 & - & - & - & 15.45 & 1 & 1 & $1.0^{\#}$ \\
 49 & Mkn1310   & 1RXP J120114.9$-$034031 & 0.019 & 1.2 & - & - & - & 15.08 & 1 & 1 & 1.0 \\
 50 & NGC4051  & 1RXP J120310.2$+$443156  & 0.002 & 19.0 & - & - & - & 12.12 & 1 & 1 & $1.0^{\#}$ \\
 51 & NGC4151   & 1RXP J121032.4$+$392418  & 0.003 & 12.5 & 3.6 & 23.5 & 0.0233$\pm$0.0049 & 11.26 & 1 & 1 & 1.0 \\
 52 & Mkn1469   &  1RXS J121607.4$+$504926 & 0.031 & 1.9 & - & - & - & 14.20 & - & 1 & 1.5 \\
 53 & NGC4235  & 1RXP J121710.1$+$071135 & 0.007 & 5.1 & - & - & - & 13.50 & 1 & 1 & 1.0 \\
 54 & Mkn766    & 1RXP J121826.4$+$294847 & 0.012 & 3.0 & - & - & - & 13.00 & 1 & 1 & $1.0^{\#}$ \\
 55 & NGC4258  & 1RXP J121856.4$+$471755 & 0.002 & 37.0 & 3.0 & 12.0 & 0.6768$\pm$0.1452 & 11.30 & 1 & 1 & 1.0 \\
 56 & NGC4278   &  1RXP J122007.2$+$291646 & 0.002 & 7.5 & 3.3 & 12.0 & 0.0713$\pm$0.0153 & 10.51 & 1 & 1 & 1.0 \\
 57 & Mkn205   &  1RXP J122144.5$+$751840 & 0.070 & 1.1 & 6.0 &  3.3 & 0.4718$\pm$0.1139 & 15.24 & 1 & 1 & 1.0 \\
 58 & Mkn50     & 1RXS J122324.4$+$024040 & 0.023 & 1.0 & - & - & - & 15.17 & - & 2 & 1.0 \\
 59 & NGC4593   &  1RXP J123939.2$-$052046 & 0.009 & 17.0 & 3.0 & 18.0 & 0.0625$\pm$0.0133 & 13.19 & 1 & 1 & 1.0 \\
 60 & NGC4594   & 1RXP J123959.2$-$113731  & 0.002 & 33.0 & - & - & - &  9.64 & 1 & 1 & 1.0 \\
 61 & NGC4639   & 1RXP J124252.2$+$131527 & 0.001 & 6.0 & - & - & - & 11.00 & 1 & 1 & 1.0 \\
 62 & IR1249-1308 & 1RXS J125212.5$-$132450 & 0.014 & 2.7 & - & - & - & 14.47 & - & 1 & $1.0^{\#}$ \\
 63 & Mkn236   & 1RXS J130021.2$+$613919  & 0.052 & 1.5 & - & - & - & 15.45 & - & 1 & 1.0 \\
 64 & Mkn783   &  1RXS J130258.8$+$162423 & 0.067 & 0.7 & - & - & - & 15.50 & - & 1 & $1.0^{\#}$ \\
 65 & NGC5033 & 1RXP J131327.7$+$363536 & 0.003 & 41.0 & - & - & - & 12.37 & 1 & 1 & 1.0 \\
 66 & Mkn1347  & 1RXS J132254.2$+$081011  & 0.050 & 1.9 & - & - & - & 14.38 & - & 2 & 1.0 \\
 67 & NGC5273   & 1RXP J134208.3$+$353919 & 0.003 & 10.0 & 3.6 & 15.5 & 0.0580$\pm$0.0124 & 13.44 & 1 & 2 & 1.0 \\
 68 & Mkn279   & 1RXH  J135303.5$+$691830  & 0.031 & 2.7 & 1.3 &  3.6 & 0.1409$\pm$0.0334 & 14.45 & 1 & 1 & 1.0 \\
 69 & Mkn662   & 1RXS J135405.7$+$232549  & 0.055 & 1.0 & - & - & - & 15.24 & - & 2 & 1.5 \\
 70 & NGC5548   & 1RXP J141759.3$+$250811 & 0.017 & 5.0 & - & - & - & 13.46 & 1 & 1 & 1.5 \\
 71 & Mkn684   & 1RXS J143104.8$+$281716 & 0.046 & 4.0 & 2.0 & 9.0 & 0.0310$\pm$0.0067 & 14.68 & 1 & 1 & $2.0^{\#}$ \\
 72 & Mkn471 & 1RXP J142255.5$+$325111 & 0.034 & 3.6 & - & - & - & 14.42 & 1 & 1 & 1.9 \\

\end{tabular}
\end{center}
\normalsize
\label{rafdats1.tab}
\end{table*}

\clearpage

\begin{table*}
\noindent {\bf Table \ref{rafdats1.tab}. \underline{continued}}
\scriptsize
\begin{center}
\begin{tabular}{rllccccccccc} \hline\hline

Nr. & \bf name & \bf ROSAT name & $z$ & $D_{\rm p}$ & $D_{\rm c}$ & $S$ & $Q$ & $V$ &\multicolumn{2}{c}{id} & Sy\\

& & & & mm & mm & mm & & mag & point.& sur. & type \\ \hline\hline

 73 & Mkn474   & 1RXP J143452.1$+$483933 & 0.041 & 1.2 & - & - & - & 15.18 & 1 & 1 & 1.0 \\
 74 & Mkn817    & 1RXP J143622.9$+$584737 & 0.033 & 2.1 & - & - & - & 13.79 & 1 & 1 & 1.0 \\
 75 & Mkn1494   & 1RXP J150139.6$+$102521 & 0.031 & 2.5 & 0.7 &  7.1 & 0.0065$\pm$0.0014 & 15.50 & 1 & - & 1.0 \\
 76 & Mkn841    & 1RXP J150401.4$+$102617 & 0.036 & 1.2 & - & - & - & 14.92 & 1 & 1 & 1.0 \\
 77 & Mkn1392   & 1RXS J150556.3$+$034212 & 0.036 & 3.1 & - & - & - & 14.14 & - & 1 & 1.5 \\
 78 & Mkn845    & 1RXS J150744.6$+$512709 & 0.042 & 2.8 & - & - & - & 14.21 & - & 1 & 1.0 \\
 79 & IR1509-2107 & 1RXP J151159.6$-$211903 & 0.044 & 0.7 & - & - & - & 14.04 & 1 & 1 & $1.0^{\#}$ \\
 80 & NGC5940  & 1RXS J153118.2$+$072713  & 0.033 & 2.6 & - & - & - & 14.20 & - & 1 & 1.0 \\
 81 & Mkn290   & 1RXP J153552.8$+$575411 & 0.029 & 1.5 & - & - & - & 14.62 & 1 & 1 & 1.0 \\
 82 & Mkn486   & 1RXP J153638.0$+$543336  & 0.039 & 1.0 & - & - & - & 14.68 & 1 & 2 & 1.0 \\
 83 & Mkn291   & 1RXP J155507.6$+$191139 & 0.035 & 1.4 & - & - & - & 15.50 & 2 & 1 & $1.0^{\#}$ \\
 84 & Mkn493    & 1RXP J155909.7$+$350154 & 0.031 & 3.1 & - & - & - & 15.06 & 1 & 1 & $1.5^{\#}$ \\
 85 & NGC6104   & 1RXS J161630.6$+$354204 & 0.028 & 6.0 & 2.0 & 16.7 & 0.0089$\pm$0.0019 & 14.00 & - & 2 & 1.0 \\
 86 & Mkn699    & 1RXP J162347.0$+$410433 & 0.034 & 0.9 & - & - & - & 15.19 & 2 & 2 & 1.2 \\
 87 & Mkn885     & 1RXS J162948.3$+$672247 & 0.026 & 2.6 & - & - & - & 14.17 & - & 1 & 1.0 \\
 88 & Mkn883  & 1RXP J162953.3$+$242640 & 0.038 & 1.2 & - & - & - & 14.43 & 1 & 1 & 1.9 \\
 89 & NGC6212  & 1RXP J164322.5$+$394823 & 0.030 & 1.9 & - & - & - & 15.00 & 1 & - & 1.0 \\
 90 & NGC6240   & 1RXP J165259.0$+$022406 & 0.024 & 5.0 & - & - & - & 13.37 & 1 & 1 & 3.0 \\
 91 & Mkn506     & 1RXP J172239.8$+$305245 & 0.043 & 2.3 & 1.7 &  3.7 & 0.1526$\pm$0.0360 & 14.55 & 1 & 1 & 1.5 \\
 92 & NGC6814   & 1RXP J194240.7$-$101928 & 0.005 & 7.8 & - & - & - & 14.37 & 1 & 1 & 1.0 \\
 93 & Mkn896   & 1RXP J204620.8$-$024848  & 0.027 & 2.3 & 0.7 &  6.0 & 0.0095$\pm$0.0021 & 14.28 & 1 & 1 & $1.0^{\#}$ \\
 94 & Mkn516 & 1RXP J215622.2$+$072213 & 0.028 & 1.7 & 1.0 & 3.6 & 0.0475$\pm$0.0113 & 15.01 & 1 & - & 1.8 \\
 95 & Mkn915   &  1RXS J223647.3$-$123228 & 0.025 & 3.4 & 2.1 &  9.8 & 0.0203$\pm$0.0044 & 14.03 & - & 2 & 1.5 \\
 96 & IR2237+0747 & 1RXH J224017.3$+$080314 & 0.025 & 3.5 & - & - & - & 14.00 & 1 & 1 & 1.0 \\
 97 & Mkn1126   & 1RXS J230048.1$-$125518  & 0.010 & 2.5 & - & - & - & 14.11 & - & 1 & $1.0^{\#}$ \\
 98 & NGC7469   & 1RXP J230315.6$+$085233 & 0.017 & 3.0 & 2.7 &  6.3 & 0.0922$\pm$0.0203 & 13.72 & 1 & 1 & 1.0 \\
 99 & Mkn315   &  1RXS J230402.8$+$223725 & 0.040 & 1.4 & 0.8 &  2.4 & 0.0857$\pm$0.0227 & 14.09 & 1 & 1 & 1.0 \\
100 & NGC7603   & 1RXP J231856.6$+$001448 & 0.029 & 4.2 & 0.8 &  4.8 & 0.0557$\pm$0.0126 & 14.21 & 1 & 1 & 1.0 \\
101 & Mkn541    & 1RXS J235602.1$+$073121 & 0.040 & 1.2 & - & - & - & 15.15 & - & 2 & 1.0 \\
102 & Mkn543    & 1RXS J000226.6$+$032105 & 0.026 & 2.0 & - & - & - & 14.09 & - & 1 & 1.0 \\ \hline\hline

\end{tabular}
\end{center}
\normalsize
\end{table*}


\begin{table*}
\caption{\bf\underline{Seyfert2 galaxies:} \rm\small{A part of the Rafanelli sample of interacting or isolated
Seyfert 2 galaxies detected by {\em ROSAT}. The table contains optical and generical properties as well as some
results from the investigations of Rafanelli et al. (1995). For the description of
columns see table \ref{rafdats1.tab}.}} 
\scriptsize
\begin{center}
\begin{tabular}{rllccccccccc} \hline\hline

Nr. & \bf name & \bf ROSAT name & $z$ & $D_{\rm p}$ & $D_{\rm c}$ & $S$ & $Q$ & $V$ &\multicolumn{2}{c}{id} & Sy\\

& & & & mm &  mm & mm & & mag & point.& sur. & type\\ \hline\hline

  1 & Mkn348 & 1RXP J004847.4$+$315716 & 0.014 & 2.0 & 0.5 & 5.0 & 0.0080$\pm$0.0018 & 14.59 & 1 & - & 2.0 \\
  2 & IR0135-1307 & 1RXP J013805.4$-$125213 & 0.041 & 1.6 & - & - & - & 15.30 & 1 & - & 2.0 \\
  3 & Mkn573 & 1RXP J014357.8$+$022050 & 0.017 & 2.3 & - & - & - & 14.07 & 1 & 1 & 2.0 \\
  4 & IR0147-0740 & 1RXP J015002.7$-$072540 & 0.017 & 1.0 & - & - & - & 15.62 & 1 & - & 2.0 \\
  5 & NGC1068 & 1RXP J024240.9$-$000042 & 0.003 & 28.0 & - & - & - & 10.83 & 1 & 1 & 2.0 \\
  6 & NGC1144 & 1RXP J025511.9$-$001042 & 0.029 & 3.2 & 1.5 & 2.4 & 0.7607$\pm$0.2015 & 14.41 & 2 & - & 2.0 \\
  7 & IR0253-1641 & 1RXS J025601.7$-$162919 & 0.033 & 1.7 & - & - & - & 15.50 & - & 1 & 2.0 \\
  8 & Mkn1066 &  & 0.012 & 4.2  & - & - & - & 13.96 & 1 & - & 2.0 \\
  9 & Mkn607 & 1RXS J032446.8$-$030256 & 0.009 & 2.5 & 6.0 & 7.5 & 0.1377$\pm$0.0300 & 14.00 & - & 1 & 2.0 \\
 10 & IR0450-0317 & 1WGA J0452.7$-$0312 & 0.016 & 2.0 & - & - & - & 15.00 & 2 & -& 2.0 \\
 11 & NGC2110 & 1RXH J055211.4$-$072725 & 0.007 & 2.5 & - & - & - & 13.51 & 1 & 1 & 2.0 \\
 12 & Mkn3 & 1RXP J061535.6$+$710209 &0.014 & 3.5 & - & - & - & 13.34 & 1 & 2 & 2.0 \\
 13 & Mkn620 &  & 0.006 & 5.0 & - & - & - & 13.54 & 1 & -& 2.0  \\
 14 & Mkn78 & 1RXP J074241.2$+$651031 & 0.038 & 1.6 & - & - & - & 14.58 & 1 & - & 2.0 \\
 15 & Mkn1210 & 1RXS J080404.6+050641 & 0.013 & 2.2 & - & - & - & 15.00 & - & 2 & 2.0 \\
 16 & NGC3081 & 1RXS J095930.3$-$224954 & 0.007 & 6.1 & 1.3 & 8.8 & 0.0328$\pm$0.0071 & 13.55 & - & 2 & 2.0 \\
 17 & Mkn720 & 1RXS J101737.6$+$065820 & 0.045 & 1.6 & - & - & - & 15.27 & - & 1 & 2.0 \\
 18 & Mkn34 &  & 0.051 & 1.4 & - & - & - & 14.65 & 1 & - & 2.0 \\
 19 & NGC3660 & 1RXS J112332.4$-$083932 & 0.011 & 6.5 & - & - & - & 14.45 & - & 1 & 2.0 \\
 20 & NGC3982 & 1RXP J115628.0$+$550731 & 0.003 & 6.9 & - & - & - & 11.70 & 1 & - & 2.0 \\
 21 & NGC4388 & 1RXP J122546.7$+$123946 & 0.008 & 16.0 & 2.3 & 39.0 & 0.0038$\pm$0.0008 & 13.90 & 1 & - & 2.0 \\
 22 & NGC4922B &  & 0.024 & 4.3 & - & - & - & 15.00 & 1 & - & 2.0 \\
 23 & NGC4941 & 1RXS J130413.2$-$053304 & 0.003 & 11.0 & - & - & - & 12.23 & - & 1 & 2.0 \\
 24 & NGC5005 & 1RXP J131056.3$+$370323 & 0.003 & 23.0 & - & - & - & 13.67 & 1 & 1 & 2.0 \\
 25 & IR1329+0216 & 1RXP J133152.2$+$020057 & 0.086 & 1.0 & - & - & - & 15.00 & 2 & 1 & 2.0 \\
 26 & NGC5252 &  & 0.022 & 4.2 & 0.7 & 8.9 & 0.0072$\pm$0.0015 & 14.21 & 1 & - & 2.0 \\
 27 & Mkn266SW & 1RXP J133818.7$+$481641 & 0.028 & 2.5 & - & - & - & 13.42 & 1 & 1 & 2.0 \\
 28 & NGC5506 & 1RXP J141315.0$-$031218 & 0.007 & 8.9 & 3.5 & 12.5 & 0.0890$\pm$0.0191 & 14.38 & 1 & 1 & 2.0 \\
 29 & Mkn670 & 1RXS J141417.5$+$264441 & 0.035 & 1.5 & - & - & - & 14.68 & 2 & 2 & 2.0 \\
 30 & Mkn673 & 1RXP J141721.4$+$265141 & 0.036 & 4.5 & - & - & - & 15.00 & 1 & - & 2.0 \\
 31 & NGC5929 & 1RXP J152607.0$+$414016 & 0.008 & 5.2 & - & - & - & 14.00 & 1 & - & 2.0 \\
 32 & NGC5953 & 1RXH  J153432.8$+$151137 & 0.007 & 4.7 & 5.5 & 3.5 & 3.0654$\pm$0.7308 & 13.10 & 1 & 1 & 2.0 \\
 33 & NGC6211 & 1RXS J164118.4$+$574601 & 0.020 & 6.5 & 2.5 & 10.0 & 0.0655$\pm$0.0141 & 14.30 & - & 2 & 2.0 \\
 34 & NGC7319 & 1RXH J223603.2$+$335833 & 0.022 & 3.5 & 4.8 & 6.2 & 0.2889$\pm$0.0638 & 13.53 & 1 & 2 & 2.0 \\
 35 & NGC7674 & 1RXP J232757.0$+$084644 & 0.029 & 4.0 & 2.0 & 11.0 & 0.0170$\pm$0.0037 & 14.36 & 1 & - & 2.0 \\
 36 & NGC7743 &  & 0.007 & 8.0 & - & - & - & 13.28 & 2 & - & 2.0 \\ \hline\hline
\end{tabular}
\end{center}
\normalsize
\label{rafdats2.tab}
\end{table*}

\clearpage


\begin{landscape}

\setcounter{table}{2}

\begin{table}
\caption{\bf\underline{Seyfert1 galaxies:} \rm\small{The table contains soft X-ray properties of the interacting
or isolated Seyfert 1 galaxies. Col(1) - object name, col(2\&3) - \ros position (pointed observations
are preferred), col(4\&5) - pointing (p = PSPC, h = HRI detector), survey (RASS
II catalogue) count rates,
col(6\&7) - exposure times, col(8\&9) - logarithmic fluxes (f = from fit, c = from count rate),
col(10\&11) - logarithmic luminosities, col(12\&13) - galactic and spectral hydrogen column densities
($\equiv$ fixed value for $N_{\rm H_{fit}} < N_{\rm H_{gal}}$ or no spectral $\Gamma$), col(14) - monochromatic flux at
1keV and col(15) - photon index. In columns 2,3,13,14 and 15 we have preferred the data from the pointed
spectral fit opposed to the survey spectral fit. Note that PSPC and HRI count rates are not comparable and
see the discussion about column densities of Seyfert 2 in Sect. \ref{interaction}.}} 
\scriptsize
\begin{center}
\begin{tabular}{lcrcccccccccccc} \hline\hline

\bf name & \multicolumn{2}{c}{ROSAT position} &\multicolumn{2}{c}{count rate} &
\multicolumn{2}{c}{$t_{\rm expo}$}
&\multicolumn{2}{c}{$\log f_{\rm X}$} & \multicolumn{2}{c}{$\log L_{\rm X}$} &
\multicolumn{2}{c}{$N_{\rm H}$} & $f_{\rm 1keV}$ & $\Gamma$  \\ 

& $\alpha_{2000}$ & $\delta_{2000}\quad$ &\multicolumn{2}{c}{$[\rm counts \; s^{-1}]$} & \multicolumn{2}{c}{[s]}
&\multicolumn{2}{c}{$[\rm erg \; cm^{-2} s^{-1}]$} & \multicolumn{2}{c}{$[\rm erg \; s^{-1}]$} & \multicolumn{2}{c}{$[10^{21}\rm cm^{-2}]$} &
$[10^{-5}\rm phot \; s^{-1}$ &  \\

& $\rm [h]\;[m]\;[s]\;\;$ & $\rm [^\circ]\;\,[']\,['']$ & pointing & survey & point. & sur. & point. & sur. & point. & sur. 
& gal & fit & $\rm cm^{-2} keV^{-1}]$ & \\ \hline


Mkn334 & 00 03 08.6 & 21 57 28 & - & 0.096$\pm$0.020 & - & 293 & - & -11.59$^c$ & - & 42.38 &
 0.443 & - & - & $\equiv\;$-2.30  \\
Mkn335 & 00 06 19.2 & 20 12 25 & 2.762$\pm$0.011$^p$ & 2.482$\pm$0.097 & 24337 & 269 & -9.95$^f$ & -9.81$^f$ & 44.14 & 44.28 &
 0.396 & 0.396$\pm$0.016 & 682.74$\pm$8.47 & -3.04$\pm$0.01 \\
Mkn1146 & 00 47 19.4 & 14 42 15 & - & 0.067$\pm$0.015 & - & 370 & - & -11.74$^c$ & - & 42.73 &
 0.460 & - & - & $\equiv\;$-2.30 \\
Mkn352 & 00 59 53.3 & 31 49 48 & 0.093$\pm$0.005$^p$ & 0.615$\pm$0.037 & 4418 & 462 & -11.63$^f$ & -10.73$^f$ & 42.01 & 42.91 &
 0.549 & 0.651$\pm$0.136 & 47.65$\pm$15.86 & -1.94$\pm$0.43 \\
Mkn1152 & 01 13 50.1 & -14 50 34 & 1.923$\pm$0.024$^p$ & 0.946$\pm$0.050 & 3219 & 412 & -10.49$^f$ & -10.98$^f$ & 44.24 &  43.74 &
 0.163 & 0.209$\pm$0.026 & 421.55$\pm$16.33  & -2.45$\pm$0.10 \\
Mkn993 & 01 25 31.4 & 32 08 00 & - & 0.083$\pm$0.017 & - & 328 & - & -11.59$^c$ & - & 42.15 & 
 0.571 & - & - & $\equiv\;$-2.30  \\
Mkn975 & 01 13 50.4 & 13 15 34 & - & 0.030$\pm$0.012 & - & 462 & - & -12.14$^c$ & - & 42.56 &
 0.390 & - & - & $\equiv\;$-2.30 \\
Mkn358 & 01 26 34.2 & 31 36 59 & - & 0.172$\pm$0.026 & - & 303 & - & -10.89$^f$ & - & 43.66 &
 0.602 & $\equiv\;$0.602 & 69.30$\pm$19.90 & -3.12$\pm$0.39 \\
Mkn359 & 01 27 32.8 & 19 10 51 & 0.822$\pm$0.016$^p$ & 0.608$\pm$0.042 & 3187 & 388 & -10.53$^f$ & -10.37$^f$ & 43.22 & 43.38 &
 0.483 & 0.543$\pm$0.068 & 356.61$\pm$16.21 & -2.51$\pm$0.13 \\
Mkn1018 & 02 06 16.0 & -00 17 32 & 0.239$\pm$0.006$^p$ & 0.081$\pm$0.017 & 6510 & 395 & -11.26$^f$ & -11.83$^c$ & 43.30 & 42.73 & 
 0.255 & 0.342$\pm$0.073 & 81.50$\pm$5.20 & -2.33$\pm$0.18 \\
Mkn590 & 02 14 33.7 & -00 46 04 & 5.225$\pm$0.039$^p$ & 2.689$\pm$0.167 & 3484 & 311 & -9.89$^f$ & -10.30$^f$ & 44.27 & 43.85 &
 0.272 & 0.317$\pm$0.020 & 1459.10$\pm$29.21 & -2.58$\pm$0.05 \\
Mkn1040& 02 28 14.6 & 31 18 38 & - & 0.342$\pm$0.033 & - & 334 & - & -11.20$^f$ & - &  42.48 & 
 0.674 & $\equiv\;$0.674 & 160.00$\pm$30.80 & -0.83$\pm$0.64 \\
Mkn1044 & 02 30 05.8 & -08 59 40 & 2.119$\pm$0.027$^p$ & 2.141$\pm$0.128 & 2836 & 278 & -10.02$^f$ & -10.12$^f$ &  43.68 &  43.58 &
 0.316 & 0.421$\pm$0.040 & 549.43$\pm$19.71 & -3.08$\pm$0.09 \\
NGC985 & 02 34 37.9 & -08 47 09 & 1.331$\pm$0.015$^p$ & 1.281$\pm$0.070 & 6191 & 267 & -10.43$^f$ & -10.41$^f$ &  44.14 &  44.16 & 
 0.283 & 0.342$\pm$0.032 & 360.98$\pm$10.91 & -2.69$\pm$0.08 \\
Mkn595 & 02 41 35.2 & 07 11 17 &  -  & 0.147$\pm$0.029 & - & 198 & - & -11.31$^c$ &  - & 42.88 & 
 0.676  & - & - & $\equiv\;$-2.30 \\
Mkn372 & 02 49 20.6 & 19 18 22 & 0.282$\pm$0.005$^p$ & 0.234$\pm$0.026 & 12675 & 368 & -10.91$^f$ & -10.76$^f$ & 43.36 & 43.52 &
 0.962 & $\equiv\;$0.962 & 173.41$\pm$4.63 & -2.37$\pm$0.07 \\
Mkn609 & 03 25 25.5 & -06 08 27 & 0.181$\pm$0.006$^p$ & 0.413$\pm$0.034 & 5801 & 400 & -11.27$^f$ & -10.81$^f$ & 43.03 & 43.50 &
 0.456 & 0.585$\pm$0.116 & 93.81$\pm$7.12 & -2.14$\pm$0.26 \\
Mkn618 & 04 36 22.1 & -10 22 31 & 0.316$\pm$0.010$^h$ & 0.579$\pm$0.042 & 3001 & 370 & -10.41$^c$ & -10.69$^f$ & 43.97 & 43.69 & 
 0.583 & $\equiv\;$0.583 & 210.00$\pm$30.70 & -2.64$\pm$0.27 \\
Mkn6 & 06 52 09.8 & 74 25 37 & - & 0.062$\pm$0.012 & - & 451 & - &  -11.69$^c$ & - & 42.15 &
 0.646 & - & - & $\equiv\;$-2.30 \\
Mkn374 & 06 59 38.5 & 54 11 36 & - & 0.775$\pm$0.050 & - & 333 & - & -10.28$^f$ & - & 44.31 &
 0.618 & $\equiv\;$0.618 & 249.00$\pm$37.60 & -3.20$\pm$0.19 \\
Mkn9 & 07 36 57.0 & 58 46 10 & - & 0.130$\pm$0.022 & - & 308 & - & -11.45$^c$ & - & 43.03 &
 0.469 & - & - & $\equiv\;$-2.30 \\
Mkn79 & 07 42 33.5 & 49 48 38 & 1.752$\pm$0.026$^p$ & 2.196$\pm$0.076 & 2690 & 383 & -9.99$^f$ & -9.94$^f$ & 43.99 & 44.03 &
 0.566 & 0.790$\pm$0.060 & 1039.90$\pm$33.85 & -2.66$\pm$0.10 \\
Mkn10 & 07 47 28.6 & 60 55 53 & 0.981$\pm$0.015$^p$ & 0.586$\pm$0.040 & 4246 & 409 & -10.49$^f$ & -10.69$^f$ & 43.75 &  43.56 & 
 0.482 & $\equiv\;$0.482 & 390.60$\pm$11.51 & -2.51$\pm$0.04 \\
Mkn382 & 07 55 25.1 & 39 11 13 & 0.122$\pm$0.006$^h$ & 0.449$\pm$0.034 & 3535 & 425& -10.65$^c$ & -10.58$^f$ & 43.71 & 43.78 &
 0.532 & $\equiv\;$0.532 & 146.00$\pm$26.90 & -3.09$\pm$0.23 \\
Mkn1218 & 08 38 10.8 & 24 53 37 & 0.473$\pm$0.013$^p$ & 0.213$\pm$0.028 & 2879 & 310 & -11.02$^f$ & -11.31$^c$ & 43.15 & 42.87 &
 0.354 & 0.449$\pm$0.045 & 235.21$\pm$49.80 & -1.50$\pm$0.22 \\ 
NGC2639 & 08 43 38.0 & 50 12 08 & 0.023$\pm$0.002$^p$ & - & 8749 & - & -12.41$^f$ & - & 40.95 & - &
 0.316 & $\equiv\;$0.316 & 6.62$\pm$1.09 & -2.16$\pm$0.24 \\
NGC2782 & 09 14 05.0 & 40 06 48 & 0.011$\pm$0.001$^h$ & - & 21715 & - & -12.19$^c$ & - & 40.90 & - &
 0.180 & - & - & $\equiv\;$-2.30 \\
Mkn704 & 09 18 26.2 & 16 18 25 & - & 0.756$\pm$0.051 & - & 321 & - & -10.61$^f$ & - & 43.61 &
 0.347 & 0.407$\pm$0.237 & 219.00$\pm$40.40 & -2.76$\pm$0.52 \\
Mkn110 & 09 25 12.8 & 52 17 13 & 6.496$\pm$0.033$^p$ & 1.691$\pm$0.063 & 6013 & 441 & -9.97$^f$ & -10.63$^f$ & 44.44 &  43.78 &
 0.156 & 0.193$\pm$0.010 & 1181.2$\pm$20.23 & -2.59$\pm$0.04 \\
Mkn705 & 09 26 03.0 & 12 44 00 & 1.001$\pm$0.014$^p$ & 1.246$\pm$0.088 & 5244 & 347 & -10.54$^f$ & -10.44$^f$ & 43.65 & 43.75 &
 0.351 & 0.398$\pm$0.042 & 342.95$\pm$11.86 & -2.53$\pm$0.09 \\
NGC2992 & 09 45 41.6 & -14 19 39 & 0.121$\pm$0.003$^p$ & 0.284$\pm$0.029 & 18586 & 394 & -11.41$^f$ & -11.08$^c$ & 41.56 & 41.89 &
 0.517 & 2.101$\pm$1.471 & 72.41$\pm$54.60 & -0.87$\pm$0.36 \\
Mkn124 & 09 48 41.6 & 50 29 26 & - & 0.028$\pm$0.010 & - & 522 & - & -12.57$^c$ & - & 42.22 &
 0.090 & - & - & $\equiv\;$-2.30 \\
Mkn1239 & 09 52 19.0 & -01 36 32 & 0.069$\pm$0.003$^p$ & 0.054$\pm$0.014 & 9043 & 418 & -10.69$^f$ & -11.89$^c$ & 43.17 & 41.96 &
 0.384 & 0.859$\pm$0.170 & 22.42$\pm$2.34 & -4.06$\pm$0.30 \\
NGC3031 & 09 55 32.8 & 69 03 54 & 0.733$\pm$0.005$^p$ & 0.998$\pm$0.088 & 28086 & 140 & -10.59$^f$ & -10.45$^f$ & 40,60 & 40,74 &
 0.427 & 0.712$\pm$0.032 & 396.29$\pm$6.19 & -2.28$\pm$0.07 \\
NGC3080 & 09 59 56.2 & 13 02 43 & 0.300$\pm$0.017$^p$ & 0.219$\pm$0.031 & 1080 & 255 & -11.10$^f$ & -11.32$^c$ & 43.28 & 43.06 &
 0.329 & 0.367$\pm$ 0.163 & 92.95$\pm$13.66 & -2.54$\pm$0.39 \\
NGC3185 & 10 17 37.8 & 21 41 24 & 0.005$\pm$0.001$^p$ & - & 4617 & - & -12.95$^f$ & - &  39.53 & - &
 0.216 & 0.279$\pm$0.894 & 1.42$\pm$1.14 & -2.47$\pm$2.64 \\
Mkn141 & 10 19 12.1 & 63 58 02 & 0.032$\pm$0.003$^h$ & 0.506$\pm$0.036 & 3589 & 498 & -11.73$^c$ & -11.15$^f$ & 42.75 &  43.33 &
 0.107 & 0.146$\pm$0.123 & 82.00$\pm$22.60 & -2.56$\pm$0.62 \\
NGC3227 & 10 23 30.2 & 19 51 50 & 0.560$\pm$0.005$^p$ & 0.099$\pm$0.020 & 19547 & 279 & -11.00$^f$ & -11.78$^c$ & 41.24 & 40.45 &
 0.222 & 0.396$\pm$0.013 & 259.21$\pm$25.76 & -1.25$\pm$0.08 \\
Mkn142 & 10 25 32.5 & 51 40 46 & 1.373$\pm$0.013$^p$ & 1.747$\pm$0.059 & 8427 & 549 & -10.41$^f$ & -10.69$^f$ & 44.21 &  43.93 &
 0.118 & 0.252$\pm$0.026 & 163.55$\pm$8.19 & -3.28$\pm$0.09 \\
Mkn634 & 10 58 01.2 & 20 29 37 & - & 0.164$\pm$0.029 & - & 236 & - & -11.62$^c$ & - & 43.32  &
 0.186 & - & - & $\equiv\;$-2.30 \\
NGC3516 & 11 06 49.0 & 72 34 07 & 4.711$\pm$0.019$^p$ & 0.140$\pm$0.020 & 13079 & 431 & -9.93$^f$ & -11.38$^f$ & 43.26 & 41.81 &
 0.355 & $\equiv\;$0.355 & 1595.40$\pm$13.08 & -2.40$\pm$0.01 \\
Mkn732 & 11 13 49.5 & 09 35 18 & - & 0.268$\pm$0.026 & - & 425 & - & -11.03$^f$ & - & 43.22 &
 0.235 & 0.438$\pm$0.407 & 98.30$\pm$28.30 & -2.64$\pm$0.96 \\
Mkn734 & 11 21 46.9 & 11 44 18 & 0.450$\pm$0.011$^p$ & 0.418$\pm$0.032 & 4011 & 439 & -10.48$^f$ & -11.02$^f$ & 44.22 & 43.67 &
 0.264 & 0.450$\pm$0.083 & 77.07$\pm$6.35 & -3.63$\pm$0.19 \\
Mkn40 & 11 25 36.6 & 54 23 03 & 0.459$\pm$0.007$^p$ & 0.101$\pm$0.020 & 8707 & 383 & -11.21$^f$ & -11.98$^c$ & 42.67 & 41.91 &
 0.106 & 0.155$\pm$0.031 & 134.72$\pm$5.77 & -1.80$\pm$0.12 \\
Mkn739A & 11 36 29.2 & 21 35 43 & 0.703$\pm$0.010$^p$ & 0.487$\pm$0.034 & 7553 & 435 & -10.50$^f$ & -11.02$^f$ & 43.74 & 43.21 &
 0.186 & 0.932$\pm$0.077 & 420.77$\pm$12.46 & -2.43$\pm$0.14 \\
Mkn744 & 11 39 42.8 & 31 54 39 & 0.391$\pm$0.012$^p$ & 0.034$\pm$0.014 & 2866 & 286 & -11.10$^f$ & -12.29$^c$ & 42.18 & 40.99 &
 0.192 & 0.652$\pm$0.283 & 206.54$\pm$61.30 & -1.00$\pm$0.18 \\
NGC3884 & 11 46 11.5 & 20 23 55 & - & 0.048$\pm$0.013 & - & 453 & - & -12.11$^c$ & - & 41.90 &
 0.214 & - & - & $\equiv\;$-2.30 \\
Mkn42 & 11 53 41.7 & 46 12 55 & 0.240$\pm$0.007$^p$ & 0.192$\pm$0.027 & 4429 & 308 & -11.24$^f$ & -11.36$^f$ & 42.81 & 42.69 &
 0.199 & 0.283$\pm$0.080 & 51.68$\pm$4.98 & -2.76$\pm$0.23 \\
Mkn1310 & 12 01 14.9 & -03 40 31 & 0.188$\pm$0.004$^p$ & 0.831$\pm$0.073 & 13559 & 339 & -11.43$^f$ & -10.82$^f$ & 42.41 &  43.02 &
 0.243 & 0.303$\pm$0.056 & 64.60$\pm$3.38 & -2.13$\pm$0.16 \\

\end{tabular}
\end{center}
\normalsize
\label{Xdats1.tab}
\end{table}

\clearpage

\begin{table}
\noindent {\bf Table \ref{Xdats1.tab}. \underline{continued}}
\scriptsize
\begin{center}
\begin{tabular}{lcrcccccccccccc} \hline\hline

\bf name & \multicolumn{2}{c}{ROSAT position} &\multicolumn{2}{c}{count rate} &
\multicolumn{2}{c}{$t_{\rm expo}$}
&\multicolumn{2}{c}{$\log f_{\rm X}$} & \multicolumn{2}{c}{$\log L_{\rm X}$} &
\multicolumn{2}{c}{$N_{\rm H}$} & $f_{\rm 1keV}$ & $\Gamma$  \\ 

& $\alpha_{2000}$ & $\delta_{2000}\quad$ &\multicolumn{2}{c}{$\rm [counts \; s^{-1}]$} & \multicolumn{2}{c}{[s]}
&\multicolumn{2}{c}{$\rm [erg \; cm^{-2} s^{-1}]$} & \multicolumn{2}{c}{$\rm [erg \; s^{-1}]$} & \multicolumn{2}{c}{$[10^{21}\rm cm^{-2}]$} &
$\rm [10^{-5}phot \; s^{-1}$ &  \\

& $\rm [h]\;[m]\;[s]\;\;$ & $\rm [^\circ]\;\,[']\,['']$ & pointing & survey & point. & sur. & point. & sur. & point. & sur. 
& gal & fit & $\rm cm^{-2} keV^{-1}]$ & \\ \hline


NGC4051 & 12 03 10.2 & 44 31 56 & 1.903$\pm$0.008$^p$ & 3.918$\pm$0.108 & 28459 & 342 & -10.42$^f$ & -10.16$^f$ & 41.46 &  41.72 &
 0.137 & 0.235$\pm$0.011 & 229.49$\pm$4.21 & -3.04$\pm$0.04 \\
NGC4151 & 12 10 32.4 & 39 24 19 & 0.600$\pm$0.004$^p$ & 0.224$\pm$0.025 & 36337 & 449 & -10.89$^f$ & -10.87$^f$ & 41.34 &  41.37 &
 0.208 & 0.282$\pm$0.019 & 162.90$\pm$3.14 & -2.46$\pm$0.06 \\
Mkn1469 & 12 16 07.4 & 50 49 26 & - & 0.079$\pm$0.017 & - & 317 & - & -11.93$^c$ & - & 42.34 &
 0.187 & - & - & $\equiv\;$-2.30 \\
NGC4235 & 12 17 10.1 & 07 11 35 & 0.184$\pm$0.004$^p$ & 0.104$\pm$0.017 & 10293 & 425 & -11.11$^f$ & -11.87$^c$ & 41.86 & 41.10 &
 0.152 & 2.091$\pm$0.263 & 154.50$\pm$96.08 & -1.96$\pm$0.60 \\
Mkn766 & 12 18 26.4 & 29 48 48 & 3.851$\pm$0.014$^p$ & 4.710$\pm$0.113 & 19856 & 377 & -10.05$^f$ & -9.80$^f$ & 43.39 &  43.64 &
 0.169 & 0.286$\pm$0.010 & 930.08$\pm$10.23 & -2.63$\pm$0.03 \\
NGC4258 & 12 18 56.4 & 47 17 56 & 0.357$\pm$0.004$^p$ & 0.195$\pm$0.021 & 25736 & 431 & -10.42$^f$ & -10.67$^f$ & 41.47 & 41.21 &
 0.120 & 0.713$\pm$0.053 & 121.04$\pm$3.96 & -3.46$\pm$0.10 \\
NGC4278 & 12 20 07.2 & 29 16 46 & 0.055$\pm$0.004$^p$ & 0.037$\pm$0.012 & 3411 & 381 & -11.41$^f$ & -12.28$^c$ & 40.47 & 39.60 &
 0.172 & 0.682$\pm$0.313 & 20.08$\pm$3.87 & -3.14$\pm$0.54 \\
Mkn205 & 12 21 44.5 & 75 18 40 & 0.780$\pm$0.010$^p$ & 0.864$\pm$0.045 & 7435 & 455 & -10.73$^f$ & -10.89$^f$ & 44.27 & 44.09 &
 0.280 & 0.350$\pm$0.038 & 265.16$\pm$8.84 & -2.37$\pm$0.09 \\
Mkn50 & 12 23 24.4 & 02 40 40 & - & 0.070$\pm$0.017 & - & 326 & - & -11.12$^f$ & - & 42.88 &
 0.176 & $\equiv\;$0.176 & 142.00$\pm$27.00 & -2.04$\pm$0.20 \\
NGC4593 & 12 39 39.2 & -05 20 46 & 1.472$\pm$0.034$^p$ & 3.432$\pm$0.170 & 1261 & 126 & -10.58$^f$ & -10.25$^f$ & 42.61 & 42.94 &
 0.228 & 0.248$\pm$0.055 & 424.54$\pm$26.58 & -2.23$\pm$0.17 \\
NGC4594 & 12 39 59.2 & -11 37 31 & 0.134$\pm$0.004$^p$ & 0.108$\pm$0.022 & 10567 & 282 & -11.44$^f$ & -11.58$^c$ & 40.44 & 40.30 &
 0.385 & 0.637$\pm$0.065 & 68.89$\pm$19.86 & -2.02$\pm$0.27 \\
NGC4639 & 12 42 52.2 & 13 15 28 & 0.035$\pm$0.002$^p$ & 0.161$\pm$0.022 & 6604 & 389 & -12.20$^f$ & -11.55$^c$ & 39.08 & 39.73 &
 0.236 & 0.312$\pm$0.207 & 10.25$\pm$1.91 & -2.22$\pm$0.68 \\
IR1249-1308 & 12 52 12.5 & -13 24 50 & - & 0.969$\pm$0.060 & - & 294 & - & -10.67$^f$ & - & 42.91 & 
 0.358 & $\equiv\;$0.358 & 265.00$\pm$39.40 & -2.50$\pm$0.17 \\
Mkn236 & 13 00 21.2 & 61 39 19 & - & 0.202$\pm$0.022 & - & 524 & - & -11.52$^f$ & - & 43.20 &
 0.183 & $\equiv\;$0.183 & 51.50$\pm$13.70 & -2.16$\pm$0.27 \\
Mkn783 & 13 02 58.8 & 16 24 23 & - & 0.288$\pm$0.034 & - & 580 & - & -11.22$^f$ & - & 43.70 &
 0.197 & 0.455$\pm$0.143 & 154.0$\pm$147.00 & -1.30$\pm$0.72 \\
NGC5033 & 13 13 27.7 & 36 35 37 & 0.277$\pm$0.007$^p$ & 0.338$\pm$0.027 & 5076 & 542 & -11.30$^f$ & -11.29$^f$ & 40.94 & 40.95 &
 0.094 & 0.250$\pm$0.067 & 90.43$\pm$6.21 & -2.10$\pm$0.24 \\ 
Mkn1347 & 13 22 54.2 & 08 10 11 & - & 0.083$\pm$0.022 & - & 270 & - & -11.89$^c$ & - & 42.80 &
 0.199 & - & - & $\equiv\;$-2.30 \\
NGC5273 & 13 42 08.3 & 35 39 20 & 0.027$\pm$0.002$^p$ & 0.021$\pm$0.008 & 5386 & 668 & -12.52$^f$ & -12.69$^c$ & 39.72 & 39.55 &
 0.093 & 0.141$\pm$0.180 & 4.94$\pm$1.52 & -2.23$\pm$0.81 \\
Mkn279 & 13 53 03.5 & 69 18 30 & 1.177$\pm$0.017$^h$ & 2.809$\pm$0.070 & 4298 & 612 & -10.20$^c$ & -10.41$^f$ & 44.07 & 43.86 &
 0.177 & 0.177$\pm$0.050 & 674.00$\pm$53.30 & -2.13$\pm$0.21 \\
Mkn662 & 13 54 05.7 & 23 25 49 & - & 0.117$\pm$0.022 & - & 326 & - & -11.78$^c$ & - & 43.00 &
 0.177 & - & - & $\equiv\;$-2.30 \\
NGC5548 & 14 17 59.3 & 25 08 11 & 5.889$\pm$0.023$^p$ & 4.954$\pm$0.105 & 10946 & 458 & -10.02$^f$ & -10.21$^f$ & 43.72 & 43.54 &
 0.193 & 0.193$\pm$0.008 & 1278.60$\pm$15.38 & -2.41$\pm$0.03 \\
Mkn684 & 14 31 04.6 & 28 17 17 & 0.245$\pm$0.009$^h$ & 0.577$\pm$0.041 & 2835 & 388 & -10.63$^c$ & -10.87$^f$ & 43.99 & 43.76 &
 0.150 & 0.267$\pm$0.170 & 105.00$\pm$25.40 & -2.86$\pm$0.53 \\
Mkn471 & 14 22 55.5 & 32 51 11 & 0.028$\pm$0.002$^p$ & 0.028$\pm$0.009 & 8971 & 562 & -12.42$^f$ & -12.52$^c$ & 41.92 & 41.83 &
 0.113 & 0.117$\pm$0.151 & 9.02$\pm$1.71 & -1.60$\pm$0.60 \\
Mkn474 & 14 34 52.1 & 48 39 34 & 0.874$\pm$0.008$^p$ & 0.248$\pm$0.020 & 12808 & 774 & -10.73$^f$ & -10.76$^f$ & 43.79 & 43.78 &
 0.201 & 0.311$\pm$0.025 & 307.97$\pm$7.32 & -2.20$\pm$0.07 \\
Mkn817 & 14 36 22.9 & 58 47 38 & 0.470$\pm$0.010$^p$ & 0.102$\pm$0.017 & 4679 & 542 & -11.22$^f$ & -11.46$^f$ & 43.11 &  42.87 &
 0.154 & $\equiv\;$0.154 & 91.61$\pm$5.26 & -2.30$\pm$0.05 \\
Mkn1494 & 15 01 39.6 & 10 25 22 & 0.006$\pm$0.001$^p$ & - & 3638  & - & -12.84$^c$ & - & 41.43 & - &
 0.224 & - & - & $\equiv\;$-2.30 \\
Mkn841 & 15 04 01.4 & 10 26 17 & 2.238$\pm$0.012$^p$ & 0.810$\pm$0.054 & 16842 & 311 & -10.35$^f$ & -10.90$^f$ & 44.06 &  43.50 &
 0.224 & 0.255$\pm$0.012 & 544.91$\pm$8.19 & -2.51$\pm$0.04 \\
Mkn1392 & 15 05 56.3 & 03 42 12 & - & 0.027$\pm$0.011 & - & 390 & - & -12.19$^c$ & - & 42.21 &
 0.374 & - & - & $\equiv\;$-2.30 \\
Mkn845 & 15 07 44.6 & 51 27 09 & - & 0.242$\pm$0.018 & - & 887 & - & -11.55$^f$ & - & 42.99 &
 0.183 & $\equiv\;$0.183 & 52.30$\pm$8.37 & -2.06$\pm$0.17 \\
IR1509-2107 & 15 11 59.7 & -21 19 00 & 0.286$\pm$0.010$^p$ & 0.067$\pm$0.015 & 2736 & 360 & -10.50$^f$ & -10.86$^f$ &  44.08 &  43.70 &
 0.792 & 3.490$\pm$0.474 & 339.41$\pm$275.20 & -2.61$\pm$1.04 \\
NGC5940 & 15 31 18.2 & 07 27 13 & - & 0.254$\pm$0.036 & - & 253 & - & -11.09$^f$ & - &  43.23 &
 0.381 & $\equiv\;$0.381 & 128.00$\pm$34.10 & -2.25$\pm$0.41 \\
Mkn290 & 15 35 52.8 & 57 54 12 & 0.592$\pm$0.010$^p$ & 0.885$\pm$0.031 & 5999 & 1061 & -10.97$^f$ & -10.82$^f$ & 43.25 & 43.39 &
 0.178 & 0.251$\pm$0.041 & 159.87$\pm$7.46 & -2.32$\pm$0.13 \\
Mkn486 & 15 36 38.0 & 54 33 37 & 0.011$\pm$0.002$^p$ & 0.003$\pm$0.002 & 2792 & 1179 & -12.98$^f$ & -13.46$^c$ & 41.50 & 41.02 &
 0.128 & $\equiv\;$0.128 & 0.89$\pm$0.87 & -2.80$\pm$0.76 \\
Mkn291 & 15 55 07.6 & 19 11 40 & 0.065$\pm$0.004$^p$ & 0.042$\pm$0.011 & 3776 & 598 & -11.30$^f$ & -12.02$^c$ & 43.09 & 42.35 &
 0.346 & 0.543$\pm$0.425 &  17.01$\pm$4.89 & -3.41$\pm$0.79 \\
Mkn493 & 15 59 09.7 & 35 01 54 & 0.379$\pm$0.007$^p$ & 0.516$\pm$0.038 & 8038 & 399 & -11.02$^f$ & -10.98$^f$ & 43.26 & 43.29 &
 0.222 & 0.295$\pm$0.050 & 77.32$\pm$4.57 & -2.84$\pm$0.14 \\
NGC6104 & 16 16 30.6 & 35 42 04 & - & 0.011$\pm$0.005 & - & 807 & - & -12.88$^c$ & - & 41.30 &
 0.133 & - & - & $\equiv\;$-2.30 \\
Mkn699 & 16 23 47.0 & 41 04 34 & 0.033$\pm$0.002$^p$ & 0.027$\pm$0.008 & 10115 & 873 & -12.46$^f$ & -12.57$^c$ & 41.90 & 41.78 &
 0.096 & 0.106$\pm$0.149 & 2.29$\pm$1.59 & -2.98$\pm$1.02 \\
Mkn885 & 16 29 48.3 & 67 22 47 & - & 0.194$\pm$0.011 & - & 1999 & - & -11.30$^f$ & - & 42.82 &
 0.378 & $\equiv\;$0.378 & 92.20$\pm$9.05 & -2.08$\pm$0.17 \\
Mkn883 & 16 29 53.3 & 24 26 40 & 0.057$\pm$0.004$^p$ & 0.041$\pm$0.011 & 3229 & 502 & -11.73$^f$ & -11.99$^c$ & 42.71 & 42.46 &
 0.403 & 1.418$\pm$0.361 & 33.65$\pm$66.16 & -2.08$\pm$1.62 \\
NGC6212 & 16 43 22.5 & 39 48 23 & 0.009$\pm$0.001$^p$ & - & 5982 & - & -13.04$^f$ & - & 41.20 & - &
 0.101 & $\equiv\;$0.101 & 1.91$\pm$0.71 & -1.88$\pm$0.40 \\
NGC6240 & 16 52 59.0 & 02 24 07 & 0.055$\pm$0.003$^p$ & 0.086$\pm$0.016 & 5232 & 507 & -10.71$^f$ & -11.59$^c$ & 43.35 & 42.46 &
 0.549 & 2.805$\pm$1.412 & 50.48$\pm$77.54 & -3.58$\pm$2.17 \\
Mkn506 & 17 22 39.8 & 30 52 46 & 0.422$\pm$0.015$^p$ & 1.526$\pm$0.047 & 1922 & 747 & -11.05$^f$ & -10.47$^f$ & 43.51 & 44.09 &
 0.328 & $\equiv\;$0.328 & 145.87$\pm$10.30 & -2.20$\pm$0.09 \\
NGC6814 & 19 42 40.7 & -10 19 29 & 0.061$\pm$0.001$^p$ & 0.363$\pm$0.035 & 38209 & 339 & -11.68$^f$ & -10.84$^f$ & 41.00 & 41.84 &
 1.296 & 1.329$\pm$0.110 &  42.37$\pm$19.16 & -1.93$\pm$0.35 \\
Mkn896 & 20 46 20.8 & -02 48 48 & 0.475$\pm$0.008$^p$ & 0.438$\pm$0.040 & 6903 & 290 & -10.38$^f$ & -10.32$^f$ & 43.78 & 43.85 &
 0.586 & $\equiv\;$0.586 & 147.68$\pm$6.04 & -3.38$\pm$0.05 \\
Mkn516 & 21 56 22.2 & 07 22 13 & 0.013$\pm$0.001$^p$ & - & 7109 & - & -11.37$^f$ & - & 42.82 & - &
 0.483 & 3.851$\pm$0.847 & 16.91$\pm$40.49 & -3.32$\pm$3.35 \\
Mkn915 & 22 36 47.3 & -12 32 28 & - & 0.123$\pm$0.026 & - & 208 & - & -11.46$^c$ & - & 42.62 &
 0.491 & - & - & $\equiv\;$-2.30 \\
IR2237+0747 & 22 40 17.2 & 08 03 17 & 0.088$\pm$0.006$^h$ & 0.484$\pm$0.051 & 2374 & 213 & -11.09$^c$ & -10.83$^f$ & 42.99 & 43.25 &
 0.659 & $\equiv\;$0.659 & 232.00$\pm$44.50 & -2.26$\pm$0.49 \\  
Mkn1126 & 23 00 48.1 & -12 55 18 & - & 0.350$\pm$0.069 & -  & 80 & - & -11.10$^c$ & - & 42.18 &
 0.352 & - & - & $\equiv\;$-2.30 \\
NGC7469 & 23 03 15.6 & 08 52 33 & 1.772$\pm$0.010$^p$ & 1.882$\pm$0.144 & 18562 & 384 & -10.21$^f$ & -10.26$^f$ & 43.53 & 43.49 &
 0.529 & 0.606$\pm$0.021 & 871.34$\pm$11.27 & -2.37$\pm$0.04 \\
Mkn315 & 23 04 02.8 & 22 37 25 & 0.022$\pm$0.002$^h$ & 0.132$\pm$0.018 & 8298 & 465 & -11.70$^c$ & -11.25$^f$ & 42.80 & 43.25 &
 0.574 & 0.578$\pm$0.676 & 75.40$\pm$28.10 & -2.42$\pm$1.68 \\
NGC7603 & 23 18 56.6 & 00 14 48 & 0.532$\pm$0.010$^p$ & 0.408$\pm$0.049 & 5425 & 189 & -10.88$^f$ & -11.08$^f$ & 43.33 & 43.13 &
 0.411 & 0.447$\pm$0.060 & 236.65$\pm$10.65 & -2.09$\pm$0.14 \\
Mkn541 & 23 56 02.1 & 07 31 21 & - & 0.049$\pm$0.013 & - & 417 & - & -11.81$^c$ & - & 42.68 &
 0.593 & - & - & $\equiv\;$-2.30 \\
Mkn543 & 00 02 26.6 & 03 21 05 & - & 0.226$\pm$0.027 & - & 402 & - & -11.02$^f$ & - & 43.10 &
 0.407 & $\equiv\;$0.407 & 88.60$\pm$21.60 & -2.73$\pm$0.29 \\ \hline\hline

\end{tabular}
\end{center}
\normalsize
\end{table}

\clearpage

\begin{table}
\caption{\bf\underline{Seyfert2 galaxies:} \rm\small{The table contains soft X-ray properties of the interacting
or isolated Seyfert 2 galaxies. For the description of columns see table \ref{Xdats1.tab}.}} 
\vspace{0.5cm}
\scriptsize
\begin{center}
\begin{tabular}{lcrcccccccccccc} \hline\hline

\bf name & \multicolumn{2}{c}{ROSAT position} &\multicolumn{2}{c}{count rate} &
\multicolumn{2}{c}{$t_{\rm expo}$}
&\multicolumn{2}{c}{$\log f_{\rm X}$} & \multicolumn{2}{c}{$\log L_{\rm X}$} &
\multicolumn{2}{c}{$N_{\rm H}$} & $f_{\rm 1keV}$ & $\Gamma$  \\ 

& $\alpha_{2000}$ & $\delta_{2000}\quad$ &\multicolumn{2}{c}{$\rm [counts \; s^{-1}]$} & \multicolumn{2}{c}{[s]}
&\multicolumn{2}{c}{$\rm [erg \; cm^{-2} s^{-1}]$} & \multicolumn{2}{c}{$\rm [erg \; s^{-1}]$} & \multicolumn{2}{c}{$[10^{21}\rm cm^{-2}]$} &
$\rm [10^{-5}phot \; s^{-1}$ &  \\

& $\rm [h]\;[m]\;[s]\;\;$ & $\rm [^\circ]\;\,[']\,['']$ & pointing & survey & point. & sur. & point. & sur. & point. & sur. 
& gal & fit & $\rm cm^{-2} keV^{-1}]$ & \\ \hline


Mkn348 & 00 48 47.4 & 31 57 16 & 0.010$\pm$0.001$^p$ & - & 23437 & - & -12.52$^f$ & - & 41.06 & - & 
 0.591 & 0.644$\pm$0.334 & 3.98$\pm$2.40 & -2.43$\pm$0.88 \\
IR0135-1307 & 01 38 05.4 & -12 52 14 & 0.010$\pm$0.001$^p$ & - & 6780 & - & -11.41$^f$ & - & 43.14 & - & 
 0.245 & 0.950$\pm$0.674 & 2.41$\pm$0.99 & -4.36$\pm$1.23 \\
Mkn573 & 01 43 57.8 & 02 20 51 & 0.076$\pm$0.002$^p$ & 0.066$\pm$0.016 & 13683 & 386 & -11.40$^f$ & -11.88$^c$ & 42.35 & 41.87 &
 0.291 & 0.430$\pm$0.107 & 15.85$\pm$1.62 & -3.32$\pm$0.24 \\
IR0147-0740 & 01 50 02.7 & -07 25 40 & 0.040$\pm$0.002$^p$ & - & 6896 & - & -11.71$^f$ & - & 42.03 & - &
 0.219 & 3.275$\pm$3.520 & 43.43$\pm$ 42.20 & -1.77$\pm$1.69 \\
NGC1068 & 02 42 40.9 & -00 00 43 & 1.957$\pm$0.040$^p$ & 1.785$\pm$0.094 & 1211 & 206 & -9.83$^f$ & -10.26$^f$ & 42.41 & 41.97 &
 0.355 & 0.524$\pm$0.069 & 478.49$\pm$28.30 & -3.44$\pm$0.14 \\
NGC1144 & 02 55 11.9 & -00 10 43 & 0.007$\pm$0.001$^p$ & - & 11855 & - & -12.07$^f$ & - & 42.16 & - & 
 0.619 & 0.879$\pm$0.689 & 2.69$\pm$0.97 & -3.46$\pm$1.14 \\
IR0253-1641 & 02 56 01.7 & -16 29 19 & - & 0.093$\pm$0.021 & - & 277 & - & -11.70$^c$ & - & 42.63 &
 0.322 & - & - & $\equiv\;$-2.30 \\
Mkn1066 & 02 59 58.7 & 36 49 17 & 0.005$\pm$0.001$^h$ & - & 15605 & - & -12.21$^c$ & - & 41.23 & - &
 1.185 & - & - & $\equiv\;$-2.30 \\ 
Mkn607 & 03 24 49.0 & -03 02 37 & - & 0.030$\pm$0.013 & - & 341 & - & -12.14$^c$ & - & 41.05 &
 0.397 & - & - & $\equiv\;$-2.30 \\
IR0450-0317 & 04 52 44.0 & -03 12 56 & 0.007$\pm$0.001$^p$ & - & 16025 & - & -12.23$^f$ & - & 41.47 & - &
 0.437 & 0.794$\pm$0.729 & 2.52$\pm$0.89 & -3.28$\pm$1.63 \\
NGC2110 & 05 52 11.6 & -07 27 25 & 0.018$\pm$0.001$^h$ & 0.037$\pm$0.010 & 31439 & 510 & -11.59$^c$ & -11.72$^c$ & 41.39 & 41.25&
 1.731 & - & - & $\equiv\;$-2.30 \\     
Mkn3 & 06 15 35.6 & 71 02 10 & 0.062$\pm$0.002$^p$ & 0.056$\pm$0.014 & 21098 & 393 & -10.63$^f$ & -11.68$^c$ & 42.95 & 41.89 &
 0.846 & 2.350$\pm$1.190 & 56.82$\pm$4.47 & -3.61$\pm$1.46 \\
Mkn620 & 06 50 09.1 & 60 50 48 & 0.001$\pm$0.000$^h$ & - & 34465 & - & -12.94$^c$ & - & 39.90 & - &
 0.731 & - & - & $\equiv\;$-2.30 \\
Mkn78 & 07 42 41.2 & 65 10 32 & 0.009$\pm$0.001$^p$ & - & 14697 & - & -11.22$^f$ & - & 43.27 & - &
 0.405 & 1.360$\pm$0.709 &  4.08$\pm$1.11 & -4.32$\pm$1.32 \\
Mkn1210 & 08 04 06.1 & 05 07 04 & - & 0.030$\pm$0.013 & - & 272 & - & -12.21$^c$ & - & 41.30 &
 0.302 & - & - & $\equiv\;$-2.30 \\ 
NGC3081 & 09 59 30.3 & -22 49 54 & - & 0.022$\pm$0.008 & - & 511 & - & -12.22$^c$ & - & 40.75 &
 0.454 & - & - & $\equiv\;$-2.30 \\ 
Mkn720 & 10 17 37.6 & 06 58 20 & - & 0.086$\pm$0.016 & - & 434 & - & -11.80$^c$ & - & 42.80 &
 0.256 & - & - & $\equiv\;$-2.30 \\ 
Mkn34 & 10 34 08.3 & 60 01 55 & 0.006$\pm$0.002$^h$ & - & 1427 & - & -12.57$^c$ & - & 42.15 & - &
 0.070 & - & - & $\equiv\;$-2.30 \\
NGC3660 & 11 23 32.4 & -08 39 32 & - & 0.061$\pm$0.016 & - & 355 & - & -11.84$^c$ & - & 41.53 &
 0.383 & - & - & $\equiv\;$-2.30 \\
NGC3982 & 11 56 28.0 & 55 07 32 & 0.015$\pm$0.002$^p$ & - & 3851 & - & -12.29$^f$ & - & 39.94 & - &
 0.122 & 0.374$\pm$0.575 & 2.24$\pm$1.30 & -3.26$\pm$1.38 \\
NGC4388 & 12 25 46.7 & 12 39 46 & 0.049$\pm$0.002$^p$ & - & 11639 & - & -11.07$^f$ & - & 42.02 & - &
 0.263 & 1.018$\pm$0.376 & 19.68$\pm$3.61 & -3.64$\pm$0.62 \\  
NGC4922B & 13 01 25.3 & 29 18 43 & 0.003$\pm$0.001$^h$ & - & 17170 & - & -12.88$^c$ & - & 41.17 & - &
 0.098 & - & - & $\equiv\;$-2.30 \\
NGC4941 & 13 04 13.2 & -05 33 04 & - & 0.035$\pm$0.015 & - & 268 & - & -12.20$^c$ & - & 40.04 &
 0.248 & - & - & $\equiv\;$-2.30 \\
NGC5005 & 13 10 56.3 & 37 03 23 & 0.089$\pm$0.003$^p$ & 0.062$\pm$0.013 & 9004 & 540 & -11.30$^f$ & -12.17$^c$ & 40.94 & 40.06 &
 0.111 & 0.558$\pm$0.155 & 27.44$\pm$2.97 & -3.12$\pm$0.31 \\
IR1329+0216 & 13 31 52.2 & 02 00 57 & 0.011$\pm$0.001$^p$ & - & 9791 & - & -12.09$^f$ & - & 43.14 & - &
 0.185 & 0.373$\pm$0.510 & 1.12$\pm$0.90 & -3.93$\pm$1.44 \\
NGC5252 & 13 38 16.0 & 04 32 31 & 0.003$\pm$0.001$^h$ & - & 3745 & - & -12.69$^c$ & - & 41.28 & - &
 0.197 & - & - & $\equiv\;$-2.30 \\ 
Mkn266SW & 13 38 18.7 & 48 16 42 & 0.039$\pm$0.002$^p$ & 0.043$\pm$0.011 & 9807 & 562 & -11.81$^f$ & -12.22$^c$ & 42.38 & 41.96 &
 0.174 & 0.410$\pm$0.177 & 9.52$\pm$1.56 & -3.04$\pm$0.40 \\
NGC5506 & 14 13 15.0 & -03 12 18 & 0.292$\pm$0.008$^p$ & 0.106$\pm$0.019 & 4380 & 366 & -10.92$^c$ & -11.58$^c$ & 42.05 & 41.40 &
 0.405 & $>$1.000 & 4070$\pm$634 & $\equiv\;$-2.30 \\
Mkn670 & 14 14 17.5 & 26 44 41 & - & 0.061$\pm$0.013 & - & 520 & - & -12.11$^c$ & - & 42.27 &
 0.149 & - & - & $\equiv\;$-2.30 \\ 
Mkn673 & 14 17 21.4 & 26 51 41 & 0.008$\pm$0.001$^p$ & - & 9205 & - & -12.32$^f$ & - & 42.10 & - &
 0.153 & 0.598$\pm$0.738 & 2.33$\pm$1.05 & -3.18$\pm$1.27 \\
NGC5929 & 15 26 07.0 & 41 40 16 & 0.013$\pm$0.001$^p$ & - & 13356 & - & -12.26$^f$ & - & 40.83 & - &
 0.208 & 0.716$\pm$0.498 & 5.68$\pm$1.16 & -2.65$\pm$0.97 \\
NGC5953 & 15 34 32.5 & 15 11 39 & 0.008$\pm$0.001$^h$ & 0.065$\pm$0.015 & 10240 & 310 & -12.25$^c$ & -11.84$^c$ & 40.73 & 41.14 &
 0.343 & - & - & $\equiv\;$-2.30 \\ 
NGC6211 & 16 41 21.5 & 57 46 24 & - & 0.008$\pm$0.004 & - & 817 & - & -12.90$^c$ & - & 40.99 &
 0.192 & - & - & $\equiv\;$-2.30 \\
NGC7319 & 22 36 03.2 & 33 58 36 & 0.003$\pm$0.001$^h$ & - & 23326 & - & -12.46$^c$ & - & 41.51 & - &
 0.773 & - & - & $\equiv\;$-2.30 \\
NGC7674 & 23 27 57.0 & 08 46 45 & 0.023$\pm$0.002$^p$ & - & 3881 & - & -10.91$^f$ & - & 43.33 & - &
 0.531 & 1.075$\pm$0.598 & 6.94$\pm$2.19 & -4.41$\pm$1.01 \\
NGC7743 & 23 44 21.4 & 09 55 57 & 0.004$\pm$0.001$^p$ & - & 18904 & - & -13.13$^f$ & - & 39.84 & - &
 0.525 & $\equiv\;$0.525 & 1.59$\pm$0.74 & -1.83$\pm$1.48 \\ \hline\hline
\end{tabular}
\end{center}
\normalsize
\label{Xdats2.tab}
\end{table}

\end{landscape}


\end{appendix}

\end{document}